\documentclass[twocolappendix]{emulateapj}
\usepackage{graphicx}
\usepackage{amssymb}
\usepackage{amsmath}
\usepackage{natbib}
\usepackage{color}
\usepackage[flushleft]{threeparttable}

\newcommand\be{\begin{equation}}
\newcommand\ee{\end{equation}}
\newcommand\ba{\begin{eqnarray}}
\newcommand\ea{\end{eqnarray}}

\begin{document}

\title{Multiple Spiral Arms in Protoplanetary Disks: Linear Theory}

\author{Ryan Miranda\altaffilmark{1,3} and Roman R. Rafikov\altaffilmark{1,2}}

\altaffiltext{1}{Institute for Advanced Study, Einstein Drive, Princeton, NJ 08540}
\altaffiltext{2}{Centre for Mathematical Sciences, Department of Applied Mathematics and Theoretical Physics, University of Cambridge, Wilberforce Road, Cambridge CB3 0WA, UK}
\altaffiltext{3}{miranda@ias.edu}


\begin{abstract}
Recent observations of protoplanetary disks, as well as simulations of planet-disk interaction, have suggested that a single planet may excite multiple spiral arms in the disk, in contrast to the previous expectations based on linear theory (predicting a one-armed density wave). We re-assess the origin of multiple arms in the framework of linear theory, by solving for the global two-dimensional response of a non-barotropic disk to an orbiting planet. We show that the formation of a secondary arm in the inner disk, at about half of the orbital radius of the planet, is a robust prediction of linear theory. This arm becomes stronger than the primary spiral at several tenths of the orbital radius of the planet. Several additional, weaker spiral arms may also form in the inner disk. On the contrary, a secondary spiral arm is unlikely to form in the outer disk. Our linear calculations, fully accounting for the global behavior of both the phases and amplitudes of perturbations, generally support the recently proposed WKB phase argument for the secondary arm origin (as caused by the intricacy of constructive interference of azimuthal harmonics of the perturbation at different radii). We provide analytical arguments showing that the process of a single spiral wake splitting up into multiple arms is a generic linear outcome of wave propagation in differentially rotating disks. It is not unique to planet-driven waves and occurs also in linear calculations of spiral wakes freely propagating with no external torques. These results are relevant for understanding formation of multiple rings and gaps in protoplanetary disks.
\end{abstract}



\section{Introduction}


The gravitational interaction of young planets with their natal disks is known to produce spiral density waves. Recent high-resolution direct imaging has revealed spiral structures in several protoplanetary disks, including MWC 758 \citep{Grady2013,Benisty2015}, HD 100453 \citep{Wagner2015,Wagner2018}, and SAO 206462 \citep{Muto2012,Garufi2013,Stolker2016,Maire2017}, which may be produced by the gravitational influence of planets or binary companions. A remarkable feature of these disks is that they display pairs of spiral arms separated by approximately $180^\circ$, which has not been expected. 

Indeed, in the conventional picture of planet-disk interactions, a planet is believed to give rise to only a single spiral arm \citep{OL02,R02}. Gravitational perturbations due to a planet excite many wave modes in the disk, each described by an azimuthal number $m$, and launched at a corresponding Lindblad resonance --- locations interior and exterior to the orbit of the planet where orbital commensurabilities occur \citep{GT79}. \citet{OL02} showed that the these modes interfere constructively, leading to a characteristic one-armed spiral pattern. In this framework, two planets would be required to produce two arms \citep{Benisty2015}. However, for pairs of arms with similar azimuthal separations to be found in several unrelated protoplanetary disks would require fortuitous configurations of the orbital phases of the planets in these systems. 

Recent three-dimensional simulations of planet-disk interactions have demonstrated that some of the observed multiple spiral features can, in fact, be produced by a single orbiting companion \citep{Zhu2015,Dong2015,Fung2015,Dong2016,Dong2017}. Notably, the spirals seen in HD 100453 were demonstrated to be consistent with the disturbances produced by the nearby M dwarf companion \citep{Dong2016_HD100453,Wagner2018}. A key finding of these numerical studies is that a single planet can produce multiple spirals, and so it is not necessary to invoke the presence of multiple planets to explain the appearance of several spiral arms. In some cases, more than two spiral arms are produced. The number of arms, as well as the azimuthal separation of the two strongest spirals, were found numerically to depend on planet mass \citep{Zhu2015,Fung2015,BZ18b}. It was also shown \citep{Bae17} that multiple arms can be related to the formation of annular gaps in mm-size dust distribution in protoplanetary disks. 

At the same time, \citet{AR18} have recently demonstrated numerically that formation of secondary spirals does not necessarily require the presence of a planet (i.e., an orbiting point mass perturber) driving density waves. In their case, density waves were driven by an imposed boundary condition at the outer edge of the simulation domain and then freely propagated inward, without angular momentum injection by external torques. Such passive propagation of the waves sufficiently far into the inner disk was found to also naturally result in the formation of a secondary arm.

Despite these numerical experiments, the origin of secondary spiral arms has remained elusive. Some nonlinear mechanisms, such as mode coupling \citep{Fung2015,Lee2016} have been proposed to explain their features. Recently, \citet{BZ18a,BZ18b} argued that the formation of multiple spirals can be explained by radially-dependent coherence of different azimuthal harmonics of the perturbations driven by a planet, essentially by an extension of the linear mode phase argument of \citet{OL02}. In their work, the phases of the multiple crests of each mode were shown to constructively interfere in different parts of the disk (at different azimuthal locations) as the wave propagates away from the perturber. The different regions of interference are identified as the primary arm, secondary arm, tertiary arm, and so on. This argument was laid out in terms of the local (WKB) approximation for mode phases and essentially ignored the behavior of the mode {\it amplitudes}. Nevertheless, these findings were corroborated by two-dimensional numerical simulations, demonstrating this idea to be a promising step towards understanding the formation of multiple spirals. An important aspect of the work of \citet{BZ18a} is that the emergence of multiple arms was understood, at least in part, within a {\it linear} framework.  

In this paper, we directly apply the linear theory of density wave evolution to self-consistently compute the full two-dimensional structure of the response of a thin, locally isothermal disk to an orbital companion. By properly accounting for the global behavior of the mode amplitudes as well as their phases (i.e., going beyond the WKB approximation), we show that multiple spiral arms are robustly formed in the inner regions of protoplanetary disks; under certain circumstances they can also appear in the outer disk. We characterize the morphology of the spirals (e.g., their amplitudes, widths, and arm-to-arm separations) and its dependence on the disk properties --- its aspect ratio, as well as profiles of the temperature and surface density. 

The plan of this paper is as follows. In Section~\ref{sec:setup}, we describe our setup and the details of our calculations. In Section~\ref{sec:results}, we present results on the formation of multiple spirals by a planet, and characterize the properties of the spirals and their dependence on the disk parameters. In Section~\ref{sec:passive} we present calculations of the passive propagation of a spiral wake in a perturber-free disk, demonstrating that the emergence of a secondary spiral is a generic property of wave propagation in differentially rotating disks. In Section~\ref{sec:analytic} we provide theoretical arguments based on linear mode phases in order to understand some key aspects of our calculations. We discuss our results in Section~\ref{sec:discussion}, and conclude with a summary of our main results in Section~\ref{sec:sum}.


\section{Problem framework}
\label{sec:setup}


We consider propagation of density waves in a two-dimensional fluid disk around a star of mass $M_*$ in the linear regime. The foundations of the mathematical framework for studying this phenomenon were laid out in \citet{GT79}, and we heavily borrow from their results. We explore both the inhomogeneous and homogeneous versions of the problem. 

In the inhomogeneous case (\S\ref{sec:results}), the wave is explicitly driven by the gravitational potential of a planet of mass $M_\mathrm{p} \ll M_*$ moving on a circular orbit with radius $r_\mathrm{p}$ and Keplerian frequency $\Omega_\mathrm{p} = (GM_*/r_\mathrm{p}^3)^{1/2}$. The torque due to the planetary gravity both excites the wave in the first place and modifies its subsequent propagation.

In the homogeneous case, the perturbation is imposed at the edge of the disk with no external torques affecting subsequent wave propagation (a setup analogous to \citealt{AR18}). This regime is studied using the same mathematical framework as the inhomogeneous case but with the planetary source terms set to zero (\S\ref{sec:passive}).


\subsection{Basic Setup}


We consider a very general disk model in which the entropy $S\propto \ln(P/\Sigma^\gamma)$ is allowed to vary with radius $r$. Here $\Sigma$ is the disk surface density, $P = \Sigma c_\mathrm{s}^2/\gamma$ is the (height-integrated) pressure, $c_s$ is an adiabatic sound speed, and $\gamma$ is the adiabatic index.
We assume that in the unperturbed disk,
\be
c_\mathrm{s}(r) = h_\mathrm{p} r_\mathrm{p}\Omega_\mathrm{p} \left(\frac{r}{r_\mathrm{p}}\right)^{-q/2},
\label{eq:cs}
\ee
where $h_\mathrm{p}$ is the disk aspect ratio, $h(r) = H/r = h_\mathrm{p}(r/r_\mathrm{p})^{(1-q)/2}$, evaluated at $r_p$, and $H = c_\mathrm{s}/\Omega$ is the pressure scale height. The exponent $q$ is the power law index of the disk temperature, which is proportional to $c_\mathrm{s}^2$. The disk surface density is
\be
\Sigma(r) = \Sigma_\mathrm{p}\left(\frac{r}{r_\mathrm{p}}\right)^{-p},
\ee
where the value of $\Sigma_\mathrm{p}$ is arbitrary. As a result of specifying $c_\mathrm{s}$ and $\Sigma$ independently, the entropy $S$ can vary through the disk.

The orbital frequency of the disk fluid is modified from the pure Keplerian value $\Omega_\mathrm{K} = (GM_*/r^3)^{1/2}$ by the pressure gradient:
\be
\Omega^2 = \Omega_\mathrm{K}^2 + \frac{1}{r\Sigma}\frac{\mathrm{d}P}{\mathrm{d}r}.
\ee
The response of the disk is sensitive to the small deviations of $\Omega$ and the radial epicyclic frequency $\kappa$, given by
\be
\kappa^2 = \frac{2\Omega}{r}\frac{\mathrm{d}}{\mathrm{d}r}(r^2\Omega),
\ee
from the Keplerian frequency $\Omega_\mathrm{K}$.


\subsection{Equations and Numerical Procedure}


The surface density perturbation produced in response to the gravitational potential of the planet (or the externally-imposed perturbation in the homogenous case) is described in polar coordinates $(r,\phi)$ by $\delta\Sigma(r,\phi)$, which we decompose into Fourier modes according to
\be
\delta\Sigma(r,\phi) = \sum_{m=1}^\infty \mathrm{Re}\left[\delta\Sigma_m(r) \mathrm{e}^{\mathrm{i}m(\phi-\phi_\mathrm{p})}\right],
\ee
where $\phi_\mathrm{p} = \Omega_\mathrm{p}t$ is the azimuthal position of the planet (in the homogeneous case, $\phi_p \rightarrow 0$). Each $\delta\Sigma_m(r)$ is a complex quantity describing the radial variation of the amplitude and phase of the mode with azimuthal number $m$. The radial velocity and azimuthal velocity perturbations $\delta u_r(r,\phi)$ and $\delta u_\phi(r,\phi)$ are similarly expressed as sums of Fourier modes $\delta u_{r,m}(r)$ and $\delta u_{\phi,m}(r)$.

\begin{figure}
\begin{center}
\includegraphics[width=0.49\textwidth,clip]{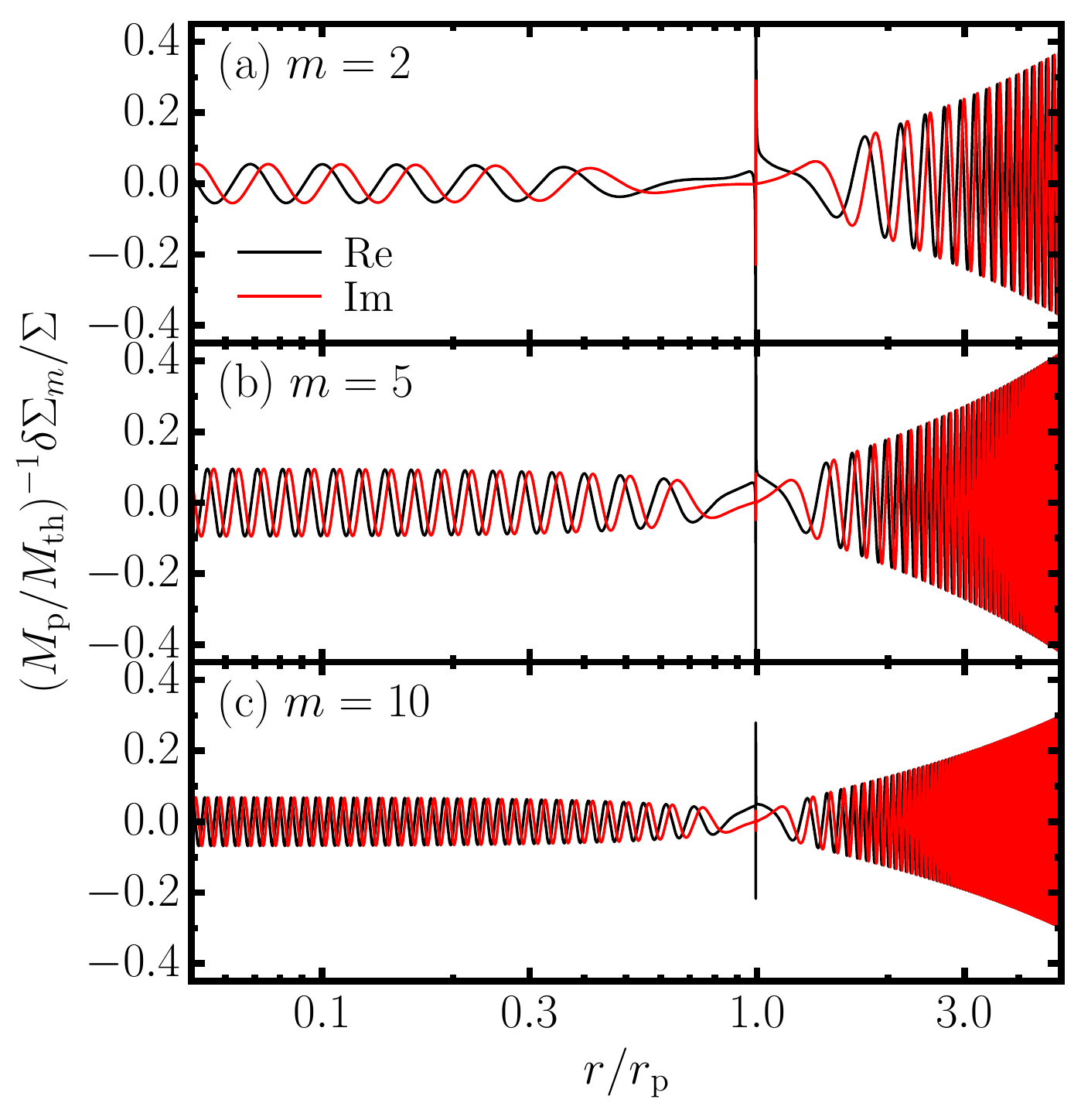}
\caption{Radial profiles of the fractional surface density perturbation for several low-order modes, for the case of the fiducial parameters, $h_\mathrm{p} = 0.1$, $q = 1$, and $p = 1$.}
\label{fig:mode_profiles}
\end{center}
\end{figure}

As a result of performing these steps, one arrives at the linear equation describing the quantity\footnote{For barotropic disks, $\delta h_m$ can be identified as the enthalpy perturbation. However, in the presence of an entropy gradient, this association no longer holds exactly. Instead, $\delta h_m$ simply serves as a convenient variable for which a master equation can be found.} $\delta h_m = \delta P_m/\Sigma$ (where $\delta P_m$ is the pressure perturbation) of the mode with azimuthal number $m$ (\citealt{Baruteau2008,Tsang2014}; here we adopt the notation of the latter):
\be
\label{eq:master_eqn}
\begin{aligned}
\left\{\frac{\mathrm{d}^2}{\mathrm{d}r^2} \right. & + \left[\frac{\mathrm{d}}{\mathrm{d}r}\ln\left(\frac{r\Sigma}{D_S}\right)\right]\frac{\mathrm{d}}{\mathrm{d}r} - \frac{2m\Omega}{r\tilde{\omega}}\left[\frac{\mathrm{d}}{\mathrm{d}r}\ln\left(\frac{\Sigma\Omega}{D_S}\right) + \frac{1}{L_S}\right] \\
& + \left. \frac{m^2}{r^2}\left(\frac{N_r^2}{\tilde{\omega}^2} - 1\right)\right\}(\delta h_m + \Phi_m) \\
& - \left[\frac{D_S}{c_\mathrm{s}^2} + \frac{1}{L_S^2} + \frac{1}{L_S}\frac{\mathrm{d}}{\mathrm{d}r}\ln\left(\frac{r\Sigma}{L_S D_S}\right) + \frac{2m\Omega}{L_S r\tilde{\omega}}\right]\delta h_m \\
& + \frac{1}{L_S} \frac{\mathrm{d}\Phi_m}{\mathrm{d}r} = 0,
\end{aligned}
\ee
where $\tilde{\omega} = m(\Omega_\mathrm{p} - \Omega)$ is the Doppler-shifted frequency of the tidal forcing due to the $m$th harmonic of the planetary potential,
\be
\label{eq:LS}
\frac{1}{L_S} = \frac{1}{\gamma}\frac{\mathrm{d} S}{\mathrm{d}r} = \frac{(\gamma-1)p-q}{\gamma r}
\ee
is the inverse length scale associated with the radial variation of entropy,
\be
\begin{aligned}
N_r^2 & = -\frac{1}{\Sigma^2}\frac{\mathrm{d}P}{\mathrm{d}r}\left(\frac{1}{c_\mathrm{s}^2}\frac{\mathrm{d}P}{\mathrm{d}r}-\frac{\mathrm{d}\Sigma}{\mathrm{d}r}\right) \\
& = \frac{(q+p)}{\gamma} \frac{c_\mathrm{s}^2}{L_S r}
\end{aligned}
\ee
is the squared Brunt--V\"{a}is\"{a}l\"{a} frequency, and $D_S = \kappa^2 - \tilde{\omega}^2 + N_r^2$. Note that in the barotropic (uniform entropy) limit, $1/L_S \rightarrow 0$ and $N_r \rightarrow 0$, equation~(\ref{eq:master_eqn}) reduces to the master equation of \citet{GT79}. The surface density perturbation $\delta\Sigma_m$ can be computed using solutions of equation~(\ref{eq:master_eqn}) according to
\be
\label{eq:dsigma_dh}
\delta\Sigma_m = \frac{\Sigma}{c_\mathrm{s}^2}\delta h_m + \frac{\mathrm{i}\Sigma}{L_S\tilde{\omega}}\delta u_{r,m},
\ee
where the radial velocity perturbation $\delta u_{r,m}$ is given in terms of $\delta h_m$ and its radial derivative $\delta h_m^\prime$ by
\be
\begin{aligned}
\delta u_{r,m} & = \frac{\mathrm{i}}{D_S}\left[\vphantom{\frac{0}{0}}\tilde{\omega}(\delta h_m^\prime + \Phi_m^\prime) \right.\\
& - \left. \frac{2m\Omega}{r}(\delta h_m + \Phi_m) - \frac{\tilde{\omega}}{L_S}\delta h_m\right].
\end{aligned}
\ee
The azimuthal velocity perturbation can be found in a similar fashion,
\be
\begin{aligned}
\delta u_{\phi,m} & = \frac{1}{D_S}\left[\frac{\kappa^2}{2\Omega}(\delta h_m^\prime + \Phi_m^\prime) \right.\\ 
& - \left. \frac{m\tilde{\omega}}{r}\left(1-\frac{N_r^2}{\tilde{\omega}^2}\right)(\delta h_m + \Phi_m) - \frac{\kappa^2}{2\Omega L_S}\delta h_m\right].
\end{aligned}
\ee

The components of the gravitational potential of the planet are
\be
\label{eq:potential_components}
\Phi_m = -\frac{GM_\mathrm{p}}{r_\mathrm{p}} b_{1/2}^{(m)}(r/r_\mathrm{p}),
\ee
where $b_{1/2}^{m}$ are (softened) Laplace coefficients,
\be
\label{eq:laplace_coef}
b_{1/2}^{(m)}(\alpha) = \frac{1}{\pi} \int_0^{2\pi} \frac{\cos(m\psi)\mathrm{d}\psi}{\left[1 - 2\alpha\cos(\psi) + \alpha^2 + \epsilon^2\right]^{1/2}}.
\ee
For the softening parameter, we choose $\epsilon = 0.6 h_\mathrm{p}$, a value which is typically used in two-dimensional numerical simulations of planet-disk interaction to mimic the vertical averaging of the planetary gravity over the disk height. In our calculations, as in \citet{BZ18a}, we ignore the ``indirect'' potential term, $\delta_{m,1}GM_\mathrm{p}r/r_\mathrm{p}^2$, which arises due to the motion of the central star around the barycenter of the star $+$ planet system. We motivate this choice in Appendix~\ref{app:solution_method}.

We solve equation (\ref{eq:master_eqn}) for a sufficient number of modes to fully capture the two-dimensional structure of the surface density perturbations. The mode solution method closely follows that of \citet{KP93} (KP93), as well as \citet{RP12} and \citet{PR12}, and is described in detail in Appendix~\ref{app:solution_method}. The solution is obtained on a logarithmic grid with $r_\mathrm{in} = 0.05 r_\mathrm{p}$ and $r_\mathrm{out} = 5.0 r_\mathrm{p}$, which, for our fiducial parameters, has $6 \times 10^4$ grid points, or a resolution of about $1300/H$. All of the modes are solved on the same grid to facilitate their synthesis, and the grid resolution is set by the tight winding of the highest $m$ modes near the grid boundaries. 

Examples of the surface density perturbation profiles $\delta\Sigma_m(r)$ for several low-order azimuthal modes are shown in Fig.~\ref{fig:mode_profiles}. Note that modes with higher $m$ are more tightly wound. Also note that here the $m = 5$ mode has the largest amplitude. This is because the response of the disk is dominated by modes with $m$ close to $m_* \approx (2h_\mathrm{p})^{-1}$ (here $h_\mathrm{p} = 0.1$ and $m_* = 5$). The dominant role of this characteristic $m$ is related to the torque cutoff phenomenon \citep{GT80}. The exact form of $m_*$ (i.e., the factor of $1/2$) is not fundamental, but this choice is supported by our numerical calculations.

Once the mode solutions are found, the two-dimensional surface density perturbation is then synthesized according to
\be
\delta\Sigma(r,\phi) = \sum_{m=1}^{m_\mathrm{max}} \mathrm{Re}\left[\delta\Sigma_m(r) \mathrm{e}^{\mathrm{i}m(\phi-\phi_\mathrm{p})}\right],
\label{eq:real}
\ee
where $m_\mathrm{max}$ is the value of $m$ necessary to achieve a converged perturbation structure; see Appendix~\ref{app:solution_method} for details. The velocity perturbations $\delta u_r(r,\phi)$ and $\delta u_\phi(r,\phi)$ are computed in the same manner. Note that the mode perturbation, $\delta\Sigma_m(r)$, is a one-dimensional, complex quantity (described by a radially-varying amplitude and phase), while the synthesized perturbation, $\delta\Sigma(r,\phi)$ is a two-dimensional, explicitly real quantity.

\begin{figure*}
\begin{center}
\includegraphics[width=0.99\textwidth,clip]{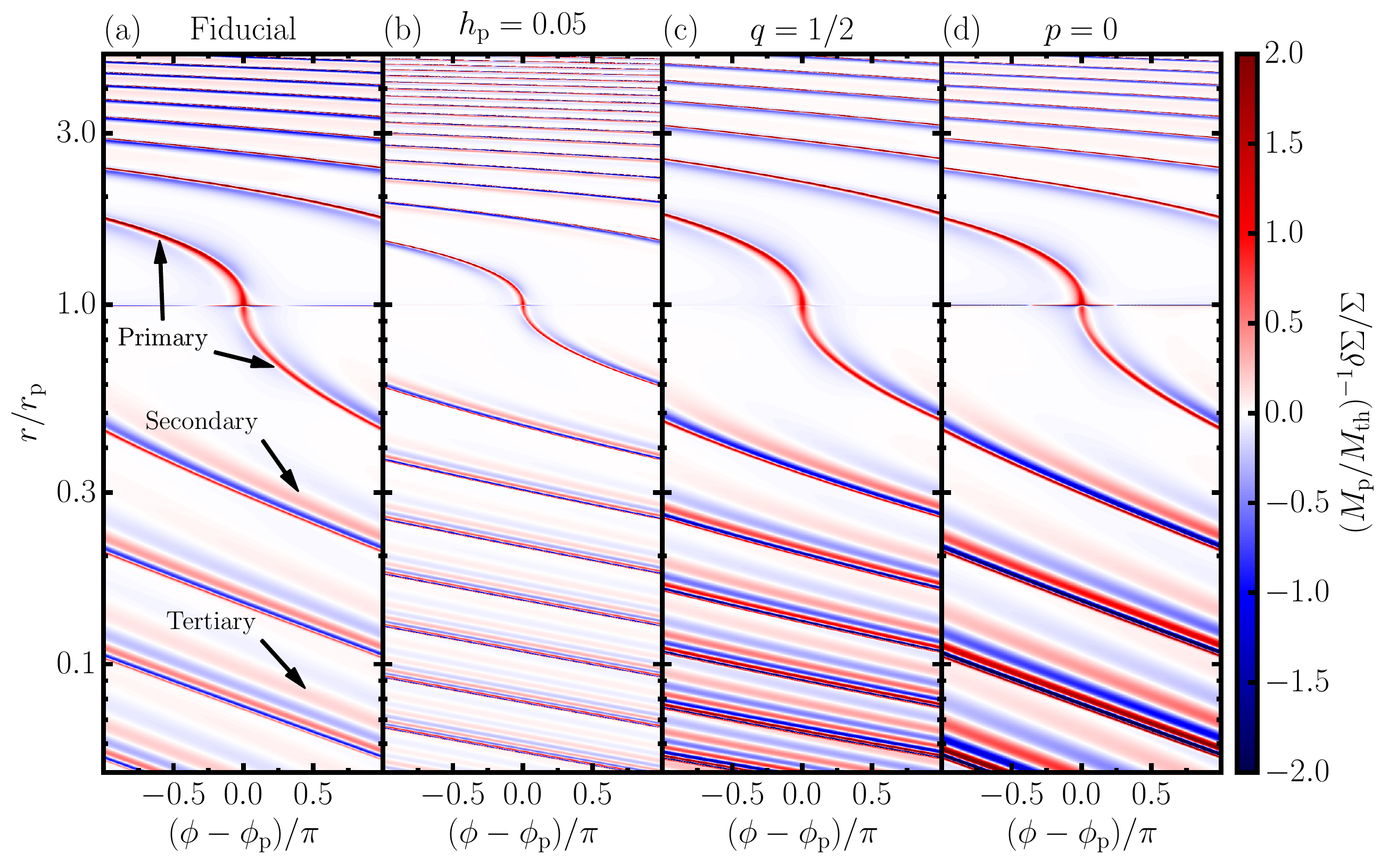}
\caption{The two-dimensional fractional surface density perturbation (scaled by the ratio of the planet mass to the thermal mass), shown in polar coordinates, for the case of the fiducial parameters ($h_\mathrm{p} = 0.1, q = 1, p = 1$; left panel), and for several other cases, demonstrating the effect of varying each parameter: the disk aspect ratio, $h_\mathrm{p}$ (middle-left panel), the temperature power-law index, $q$ (middle-right panel), and the surface density power-law index, $p$ (right panel). The positions of the primary arm, secondary arm, and tertiary arm are indicated in the leftmost panel.}
\label{fig:sigma2d}
\end{center}
\end{figure*}


\subsection{Parameters}


The results of our calculations are fully determined by four dimensionless parameters: $h_\mathrm{p}$, the disk aspect ratio at the orbital radius of the planet, $q$, the power law index of the disk temperature, $p$, the power law index of the disk surface density, and the adiabatic index $\gamma$. For the fiducial parameters, we choose $h_\mathrm{p} = 0.1$, $q = 1$ (corresponding to a constant disk aspect ratio $h$), $p = 1$, and $\gamma = 7/5$. We find that our results are almost completely insensitive to the value of $\gamma$ (see \S \ref{subsec:gamma}), and so unless otherwise stated, we keep its value fixed. We have explored the parameter space by performing calculations for which two of the three remaining parameters are fixed at their fiducial values and the third is varied over a plausible range of values for protoplanetary disks: $0.05 < h_\mathrm{p} < 0.15$, $0 < q < 1$, $0 < p < 3/2$. We find that the results are not sensitive to $p$, and so we primarily focus our analysis and discussion on the effects of varying $h_\mathrm{p}$ and $q$.

\begin{figure*}
\begin{center}
\includegraphics[width=0.99\textwidth,clip]{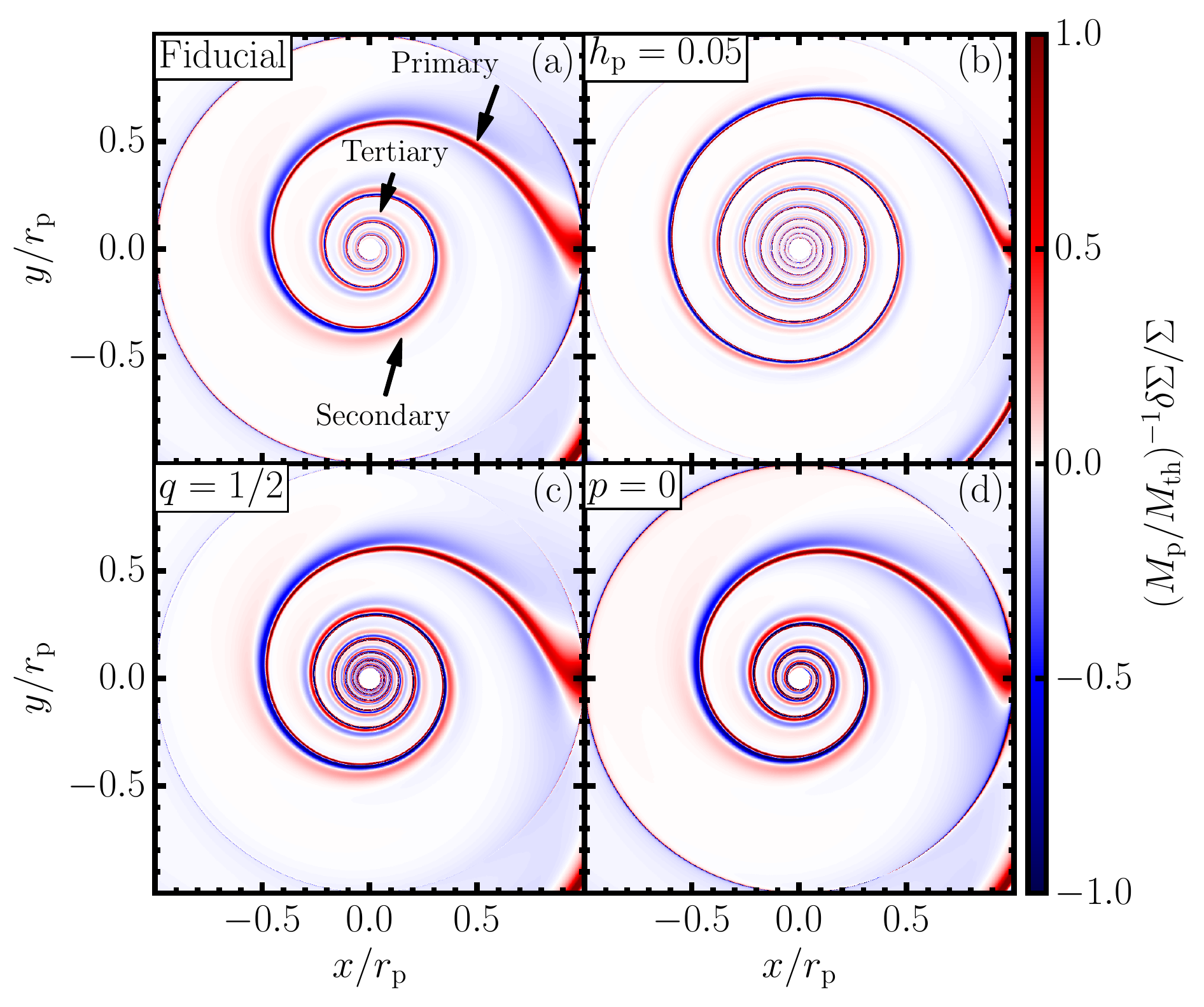}
\caption{The two-dimensional fractional surface density perturbation for the fiducial paramaters and several cases with varied parameters, as in Figure~\ref{fig:sigma2d}, but shown here in Cartesian coordinates, and focusing only on the spiral structure in the inner disk, interior to the orbit of the planet. The positions of the primary, secondary, and tertiary arms are indicated in the top-left panel.}
\label{fig:sigma2d_cartesian}
\end{center}
\end{figure*}


\section{Results for Planet-Driven Spirals}
\label{sec:results}


We will start presentation of our results with the case of a spiral pattern driven by the gravity of an embedded planet (forced or inhomogeneous case). To better highlight new findings, we start by outlining the existing picture of wave propagation in disks in \S \ref{sect:expectations}. We then describe general properties of the linear planet-driven density waves found in our linear calculations (\S \ref{sect:gen_res}) and provide a comparison with the results of direct numerical simulations (\S \ref{sect:num_val}). These preliminaries form a basis for subsequent in-depth discussion of the properties of multiple spiral arms emerging in our calculations in \S \ref{subsec:spiral_characterization}.


\subsection{Expectations Based on Simple Linear Theory}
\label{sect:expectations}


Existing linear calculations of the density wave propagation in disks provide some guidance on the expected outcome of our present calculation. Specializing to the inhomogeneous case, the linear response of the disk is expected to take a form of a one-armed spiral density wake \citep{R02,OL02}. For $|r - r_\mathrm{p}| \gg H_\mathrm{p}$, the position of the wake is given approximately by\footnote{Note that analogous expression in \citet{R02} has a different sign of the second term.}
\be
\label{eq:phi_lin}
\phi_\mathrm{lin} = \phi_\mathrm{p} + \mathrm{sgn}(r-r_\mathrm{p})\int_{r_\mathrm{p}}^r\frac{\Omega(r^\prime)-\Omega_\mathrm{p}}{c_\mathrm{s}(r^\prime)}\mathrm{d}r^\prime,
\ee
and its amplitude (peak height) is given by, to within a constant factor of order unity \citep{R02},
\be
\label{eq:amp_lin}
\begin{aligned}
\frac{\delta\Sigma_\mathrm{lin}}{\Sigma(r)} & = \frac{M_\mathrm{p}}{M_\mathrm{th}}  \\
& \times \left[\frac{\Omega(r)-\Omega_\mathrm{p}}{\Omega_\mathrm{p}}
\frac{\Sigma_\mathrm{p}}{\Sigma(r)}
\frac{r_\mathrm{p}}{r}\right]^{1/2}
\left[\frac{c_\mathrm{s}(r)}{c_\mathrm{s}(r_\mathrm{p})}\right]^{-3/2}.
\end{aligned}
\ee
Here
\be
M_\mathrm{th} = h_\mathrm{p}^3 M_*
\ee
is the thermal mass. The first (constant) factor on the right-hand side of Equation (\ref{eq:amp_lin}) describes the initial amplitude of the density wake formed within a few scale heights of the planet (in the homogeneous case considered in \S \ref{sec:passive}, $r_p$ is replaced with the radius at which the perturbation is imposed, and $M_\mathrm{p}/M_\mathrm{th}$ with an arbitrary constant), while the other factors indicate how the amplitude varies as the wake propagates away from the planet. The radial scaling is dictated by the conservation of angular momentum flux (AMF),
\be
F_J(r) = r^2 \Sigma(r) \oint \delta u_r(r,\phi) \delta u_\phi(r,\phi) \mathrm{d}\phi, 
\ee
which in the absence of explicit dissipation (linear or nonlinear) must be constant far from the planet, outside the wave excitation region \citep{GR01,R02}. The characteristic scale of $F_J$, resulting from the sum of the one-sided Lindblad torques, is \citep{GT80,Ward1997}
\be
\label{eq:FJ0}
F_{J,0} = \left(\frac{M_\mathrm{p}}{M_*}\right)^2 h_\mathrm{p}^{-3} \Sigma_\mathrm{p} r_\mathrm{p}^4 \Omega_\mathrm{p}^2.
\ee
These known linear results will be used as a reference for comparison in our current calculations.

\begin{figure*}
\begin{center}
\includegraphics[width=0.99\textwidth,clip]{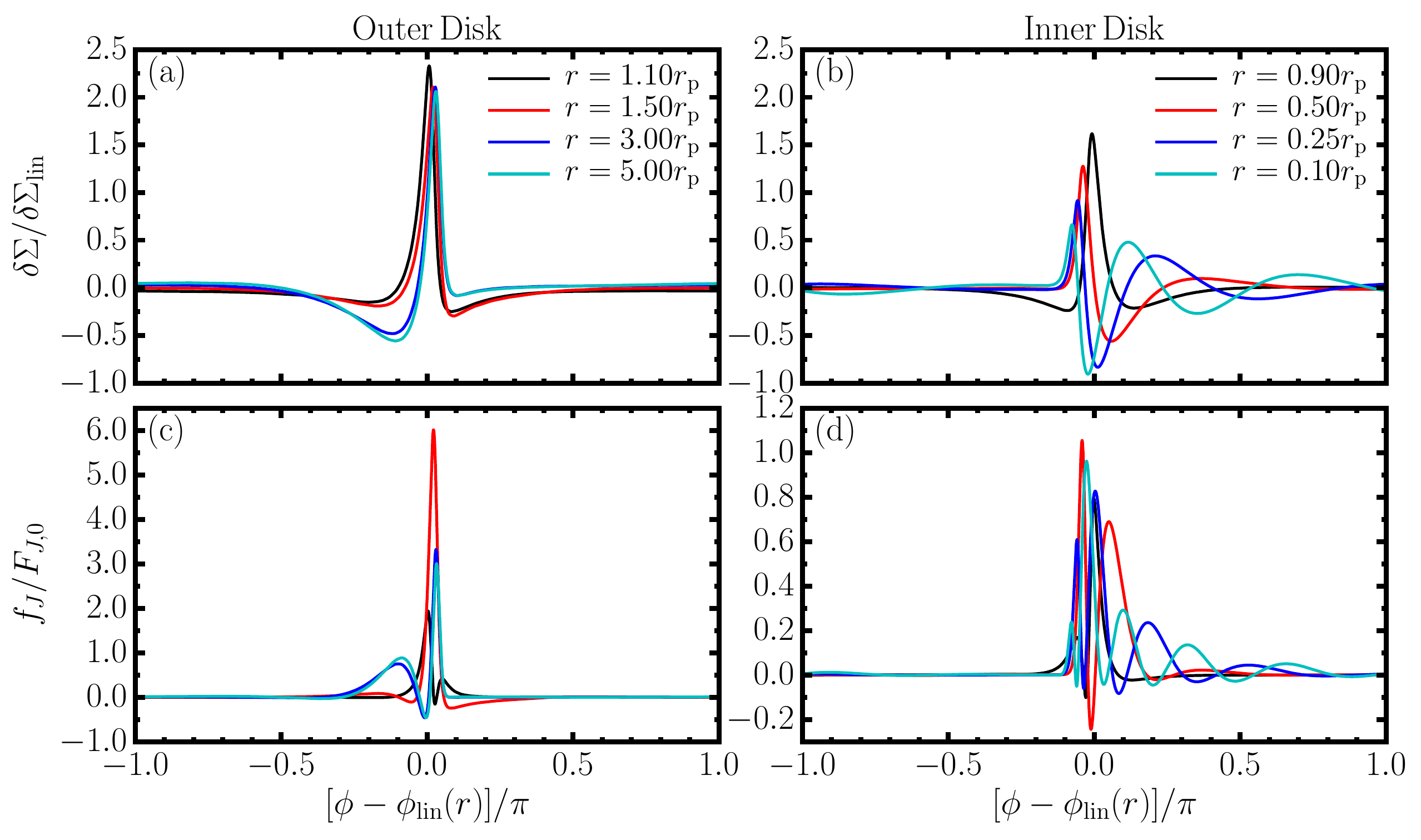}
\caption{\textit{Top panels}: profiles of the surface density perturbation $\delta\Sigma$, scaled by $\delta\Sigma_\mathrm{lin}$ (see equation~\ref{eq:amp_lin}) to facilitate comparison between the profiles at different radii. \textit{Bottom panels}: profiles of the azimuthal distribution of angular momentum flux $f_J$, in terms of the characteristic angular momentum flux $F_{J,0}$ (equation \ref{eq:FJ0}). Profiles are shown for several different radii in the outer disk (left panels) and the inner disk (right panels), for the fiducial disk parameters.}
\label{fig:wake_profiles}
\end{center}
\end{figure*}


\subsection{Results of Linear Calculations: General Properties}
\label{sect:gen_res}


In Figure~\ref{fig:sigma2d} we show two-dimensional maps of the perturbed surface density (in polar coordinates $r,\phi$) resulting from our typical linear calculations. The case with the fiducial parameters is shown (Fig.~\ref{fig:sigma2d}a), as well as several other cases, discussed in \S \ref{sect:par_dep}, in which each of the three parameters (indicated at the top of the panel) is varied from its fiducial value.

The same maps of the perturbed surface density are also shown in Cartesian coordinates in Figure~\ref{fig:sigma2d_cartesian}, in order to demonstrate the true geometry of the spirals as seen by an observer sensitive to the gas surface density perturbations. Note that in Figure~\ref{fig:sigma2d_cartesian}, only the part of the disk interior to the orbit of the planet is shown, and the range of color scale for the perturbations is smaller than in Figure~\ref{fig:sigma2d}, in order to focus on the detail of the perturbations in the inner disk.\footnote{We point out that radially narrow features seen at $r \approx r_\mathrm{p}$ in Figs.~\ref{fig:sigma2d}--\ref{fig:sigma2d_cartesian} are perturbations in the corotation region which result from the presence of a radial entropy gradient (i.e., due to terms depending on $1/L_S$ in equation~\ref{eq:master_eqn}). As they are very localized near the orbit of the planet and have no effect on the global spiral structure of the disk, we do not examine these features in detail in this work.}

Also, to highlight the details of the spiral arm evolution, in Figure~\ref{fig:wake_profiles}a,b we display azimuthal profiles of $\delta\Sigma$ at a fixed radius $r$, normalized by the linear wake amplitude $\delta\Sigma_{\rm lin}$ given by equation~(\ref{eq:amp_lin}).  These profiles can be thought of as horizontal cuts in Figure \ref{fig:sigma2d}.

Generically, in the outer disk ($r > r_\mathrm{p}$), a single strong spiral arm, i.e., a narrow structure with a $\delta\Sigma > 0$ peak is present. Its azimuthal profile remains relatively unchanged as the arm winds up with increasing $r$, see Figure~\ref{fig:wake_profiles}a. It is accompanied by a comparatively wider surface density trough, with $\delta\Sigma < 0$, which trails behind the arm (in the sense of the Keplerian rotation of the disk). The case with a smaller disk aspect ratio, $h_\mathrm{p} = 0.05$, represents an exception, as it develops a weak additional spiral arm beyond $r \approx 3 r_\mathrm{p}$, although this is not clearly discernible in Figure~\ref{fig:sigma2d} (see \S \ref{subsec:spiral_characterization} for details).

In the inner disk ($r < r_\mathrm{p}$), there is a single spiral arm (the ``primary spiral'') near the planet, accompanied by a trough which leads the arm. However, farther from the planet, the structure of the density wake deviates from the simple behavior seen in the outer disk. For the fiducial parameters, the presence of second peak with $\delta\Sigma > 0$ (the ``secondary spiral arm'') becomes apparent at $r \approx 0.5 r_\mathrm{p}$, accompanied by a deepening of the initial trough, see Figure \ref{fig:wake_profiles}b. Towards smaller radii, the strength of the secondary spiral increases, and it also becomes accompanied by a leading trough. At even smaller radii ($r \lesssim 0.1 r_\mathrm{p}$), a third arm (the ``tertiary spiral'') forms. These multiple spirals are robustly present for a variety of disk parameters. Note that the additional spirals (secondary, tertiary, and so on) are always located {\it ahead of} the primary spiral.

In Figure \ref{fig:wake_profiles}c,d we also show the azimuthal distribution of angular momentum flux,
\be
\label{eq:dFJdphi}
f_J(r,\phi) = r^2 \Sigma(r) \delta u_r(r,\phi) \delta u_\phi(r,\phi),
\ee
at different radii. This quantity is related to the angular momentum flux according to
\be
F_J(r) = \oint f_J(r,\phi) \mathrm{d}\phi.
\ee
The evolution of the azimuthal profile of $f_J$ with $r$ allows us to trace the exchange of angular momentum flux between the multiple arms as the wake propagates away from the planet.

The AMF is peaked near the locations of both the peaks ($\delta\Sigma > 0$) and the troughs ($\delta\Sigma < 0$) of the surface density perturbation, since the angular momentum flux is approximately proportional to the square of $\delta\Sigma$ far from the planet \citep{R02},
\be
\label{eq:dFJdphi_approx}
f_J \approx \frac{r c_\mathrm{s}^3}{|\Omega-\Omega_\mathrm{p}|\Sigma} (\delta\Sigma)^2.
\ee
In particular, in the outer disk, for $r/r_\mathrm{p} \gtrsim$ a few, the trough trailing the primary arm of the surface density perturbation carries a significant fraction of the AMF.

We provide more in depth discussion of the different features of the density wake in \S \ref{subsec:spiral_characterization}.


\subsection{Dependence on Adiabatic Index}
\label{subsec:gamma}

\begin{figure}
\begin{center}
\includegraphics[width=0.49\textwidth,clip]{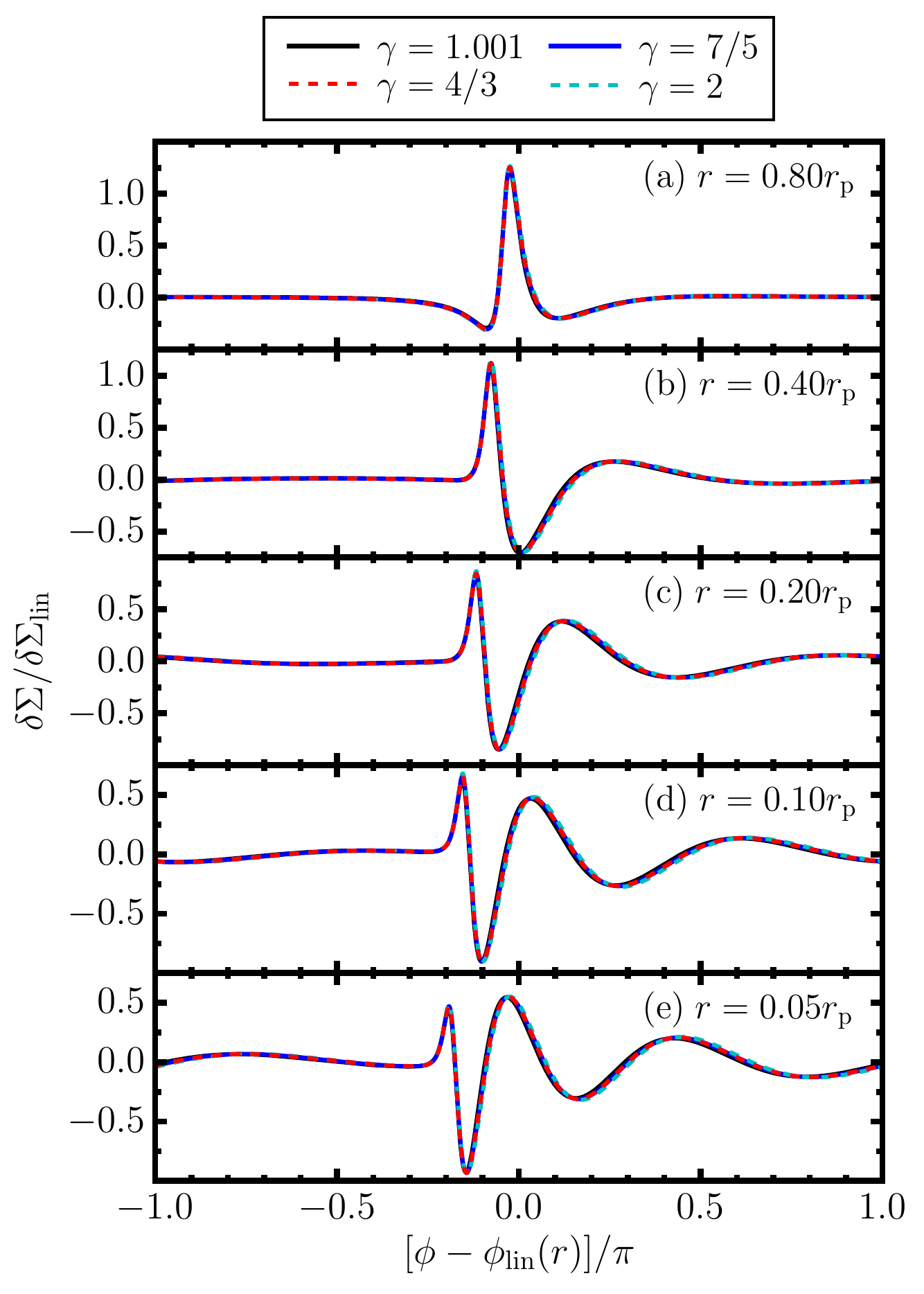}
\caption{Surface density perturbation profiles in the inner disk for the fiducial disk parameters, $h_\mathrm{p} = 0.1$, $q = p = 1$, and different values of the adiabatic index $\gamma$.}
\label{fig:wake_profiles_gamma}
\end{center}
\end{figure}

In the calculations presented in this section, we have adopted $\gamma = 7/5$ as the fiducial adiabatic index. We have also carried out calculations for $h_\mathrm{p} = 0.1$, $q = 1$, and $p = 1$ with different values of $\gamma$, the results of which are shown in Figure~\ref{fig:wake_profiles_gamma}. Here azimuthal profiles of $\delta\Sigma$ are shown at different radii for several values of $\gamma$ in addition to $\gamma=7/5$. The case $\gamma = 1.001$ corresponds to an almost (locally) isothermal disk. The case $\gamma = 4/3$ represents the effective two-dimensional adiabatic index corresponding to a three-dimensional adiabatic index $\Gamma = 7/5$. These different adiabatic indices are related according to $\gamma = (3\Gamma-1)/(\Gamma+1)$ \citep{Goldreich1986,Ostriker1992} in the low-frequency limit, $\tilde{\omega} \ll \Omega$ (valid for $r \approx r_\mathrm{p}$). For $\gamma = 2$, the disk has a uniform entropy profile (see equation~\ref{eq:LS}) and we recover the barotropic limit. 

The surface density profiles shown in Figure~\ref{fig:wake_profiles_gamma} demonstrate that our results are remarkably insensitive to the value of $\gamma$\footnote{The perturbation structure in the immediate vicinity of $r_\mathrm{p}$ (within about $0.1H$) has a strong dependence on $\gamma$, but this is not important for our present study of spiral arms far from the planet.}. Mathematically, this is because $\gamma$-dependent contributions appear in equation (\ref{eq:master_eqn}) only through terms varying on scales of $\sim r_p$. In \S \ref{sec:analytic}, we show that the formation of multiple spirals is well-described by considering the phases of different modes in the local (WKB) limit, in which the globally-varying terms in equation~(\ref{eq:master_eqn}) with explicit dependence on $\gamma$ are unimportant. Therefore, it is sufficient to only consider the fiducial $\gamma = 7/5$ in our subsequent discussion.

\subsection{Numerical Validation}
\label{sect:num_val}


To validate and understand the limitations of our semi-analytical linear calculations, we have also carried out a set of direct numerical simulations of planet-disk interaction in the low mass regime using {\sc fargo3d} \citep{FARGO3D}. These simulations will be discussed in detail in a future work. Here we describe only the basic setup as well as the results for the fiducial disk parameters, $h_\mathrm{p} = 0.1$, $q = 1$, and $p = 1$.

\begin{figure}
\begin{center}
\includegraphics[width=0.49\textwidth,clip]{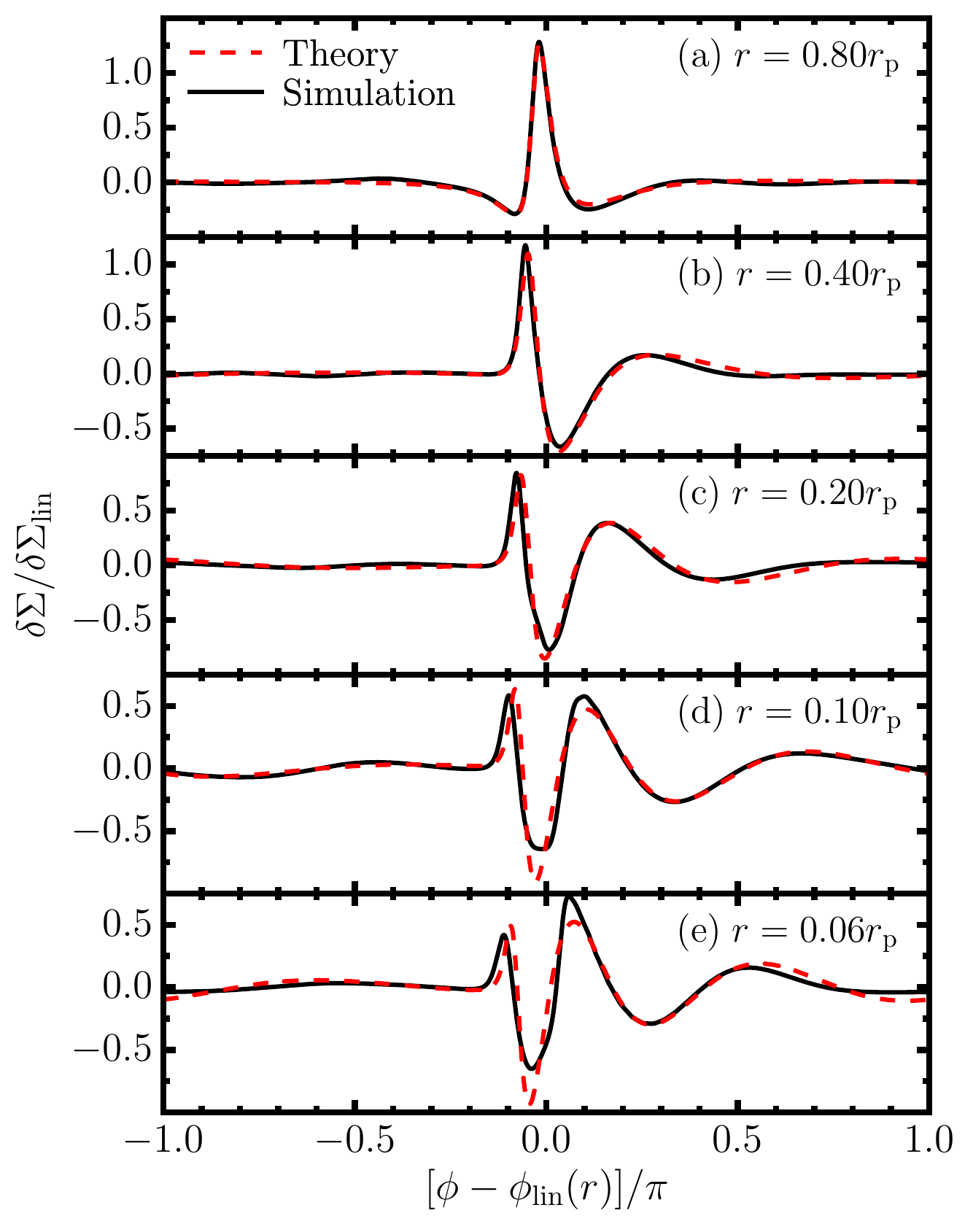}
\caption{Profiles of the surface density perturbation at several radii in the inner disk, comparing the results of our numerical simulation (solid curves) and the results of our linear calculation (dashed curves) for the fiducial parameters (with $\gamma = 1.001$).}
\label{fig:wake_profiles_sim}
\end{center}
\end{figure}

We choose a planet mass $M_\mathrm{p} = 10^{-5} M_*$, or $0.01 M_\mathrm{th}$. Since $M_\mathrm{p} \ll M_\mathrm{th}$, the linear regime is appropriate for describing the disk response, except far ($\gtrsim 6 H_\mathrm{p}$) from the planet, where nonlinear effects (fully captured in simulations)---wake evolution into a shock and subsequent dissipation---become non-negligible \citep{GR01}. A softening length of $0.6 H_\mathrm{p}$ is applied to the potential of the planet. We use a logarithmically-spaced radial grid extending from $r_\mathrm{in} = 0.05 r_\mathrm{p}$ to $r_\mathrm{out} = 5.0 r_\mathrm{p}$ with $N_r = 4505$ grid cells, and a uniformly-spaced azimuthal grid with $N_\phi = 6144$ grid cells. The resulting grid cells have a roughly square shape with a resolution of $98$ cells per scale height. Wave damping zones (e.g., \citealt{deValBorro2006}) are implemented for $r < 0.06 r_\mathrm{p}$ and $r > 4.5 r_\mathrm{p}$ in order to minimize wave reflection at the boundaries (note that direct comparison with the results of our linear calculations is not possible in the damping zones). 

We adopt an ideal equation of state with $\gamma = 1.001$. This value of $\gamma$ close to unity is chosen in order to avoid any significant heating of the disk due to wave dissipation during our simulations. Note that although this setup resembles that of a locally isothermal disk, the two cases are physically distinct. In locally isothermal disks, non-axisymmetric perturbations are known to exchange angular momentum with the background flow \citep{Lin2011,Lin2015}, which can lead to anomalous results.\footnote{We look into this issue in more detail in Miranda \& Rafikov (in prep.).} No explicit viscosity is included in the simulation.

We compare the results of the numerical simulation after $10$ orbits (when steady state is reached) with our linear calculations (using $\gamma = 1.001$ in order to facilitate direct comparison) in Figure~\ref{fig:wake_profiles_sim}. The profiles of the surface density perturbation $\delta\Sigma$ are shown at several different disk radii in the inner disk, highlighting the multiple spiral arm structure. The profiles are rotated by $\phi_\mathrm{lin}(r)$ and scaled by $\delta\Sigma_\mathrm{lin}(r)$, as in Figure~\ref{fig:wake_profiles}. 

The numerical results exhibit very good agreement with the linear prediction. The positions of the primary, secondary, and tertiary arms are closely reproduced, differing by $\lesssim 0.05$ radians from the linear prediction even at the smallest radii. In particular, the initial secondary-to-primary arm separation is $\approx 60^\circ$, and this separation decreases towards smaller radii, in agreement with our calculations (see \S \ref{subsec:spiral_characterization}). This is a unique feature of the secondary spiral in the linear regime, which should be contrasted with the $180^\circ$ separation which arises when $M_\mathrm{p} > M_\mathrm{th}$ (e.g., \citealt{Zhu2015}). The amplitudes of the arms and troughs in the numerical simulation also show good agreement with the linear prediction, exhibiting essentially negligible differences, except for at small radii ($r \lesssim 0.1 r_\mathrm{p}$), where differences of $\approx 20\%$ arise due to nonlinear effects (especially in the amplitude of the trough between the primary and secondary spirals). Notably, the amplitude of the secondary arm overtaking that of the primary for $r \lesssim 0.1 r_\mathrm{p}$ is well reproduced in the simulation. The agreement between the numerical simulation and linear theory supports full applicability of our results to low-mass planets.

We note that no shocks are present in Figure~\ref{fig:wake_profiles_sim}, in contradiction with the results of \citet{R02}, which predict shock formation at about $0.3 r_\mathrm{p}$. However, the calculations of \citet{R02} were carried out for the case of a single spiral arm propagating in a self-similar fashion. In our calculations, there is an exchange of angular momentum flux from the primary spiral arm to the secondary spiral arm, which modifies this picture. Evidently this exchange lowers the amplitude of the primary arm and suppresses shock formation for the planet mass ($M_\mathrm{p} = 0.01 M_\mathrm{th}$) that we have considered here.


\subsection{Structure of the Multiple Spirals}
\label{subsec:spiral_characterization}


\begin{figure*}
\begin{center}
\includegraphics[width=0.99\textwidth,clip]{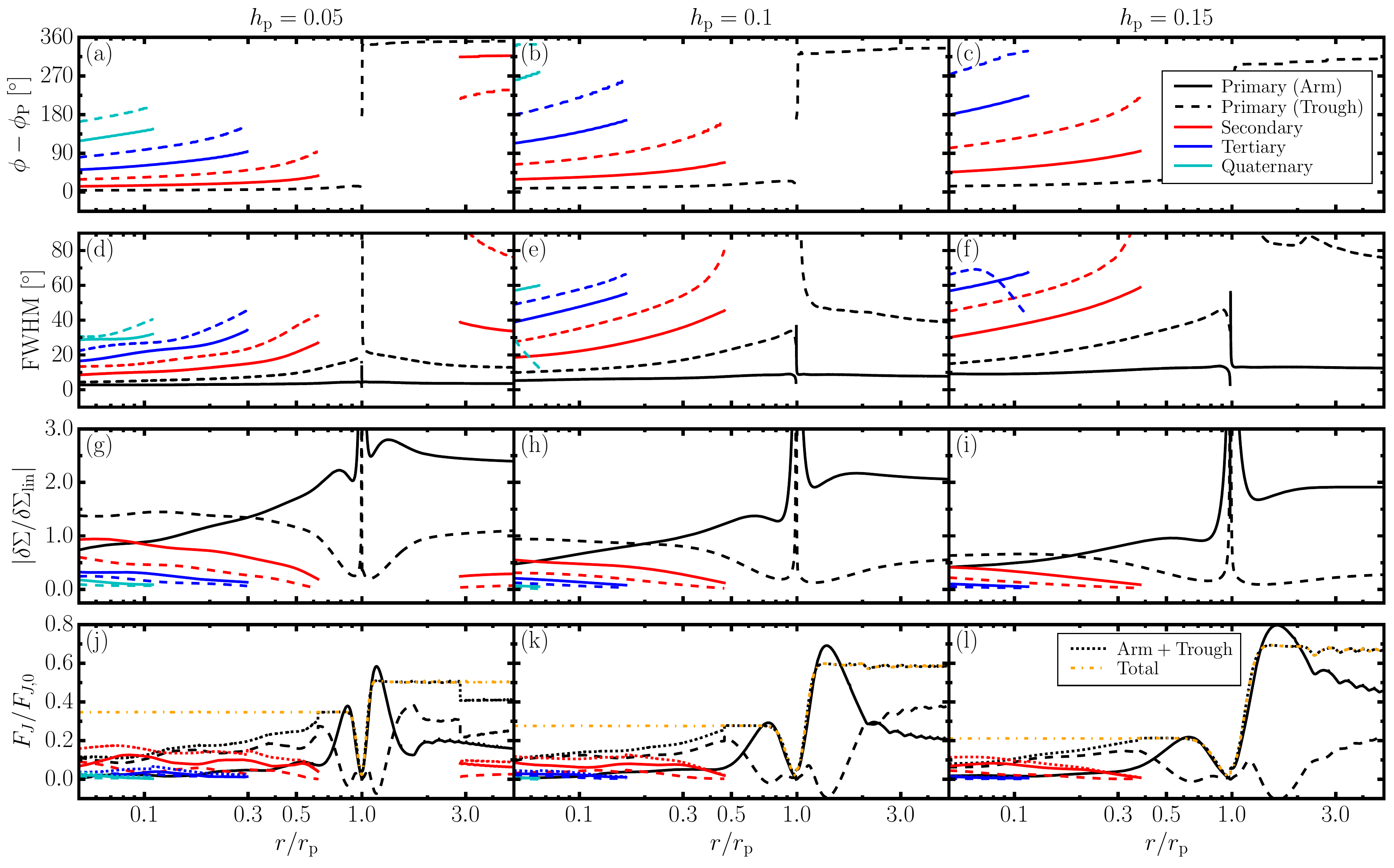}
\caption{Properties of the multiple spiral arms (solid curves) and their associated troughs (dashed curves) as functions of radius, for different values of the disk aspect ratio near the planet, $h_\mathrm{p}$ (left, right, and middle panels), and with fixed values of the temperature and surface density power law indices $q = 1$ and $p = 1$. \textit{Top panels}: Positions of the arms and troughs relative to the primary arm. \textit{Top middle panels}: The full-width half maxima of the arms and troughs. \textit{Bottom middle panels}: The absolute values of the surface density perturbations associated with the arms and troughs, scaled by $\delta\Sigma_\mathrm{lin}$ (equation \ref{eq:amp_lin}). \textit{Bottom panels}: The angular momentum flux $F_\mathrm{J}$ associated with each arm or trough, as well as the sum of the angular momentum flux of each arm and its associated trough (dotted curves), and the total angular momentum flux (sum over all of the arms and troughs; dot-dashed curves).}
\label{fig:arm_profiles_h}
\end{center}
\end{figure*}

\begin{figure*}
\begin{center}
\includegraphics[width=0.66\textwidth,clip]{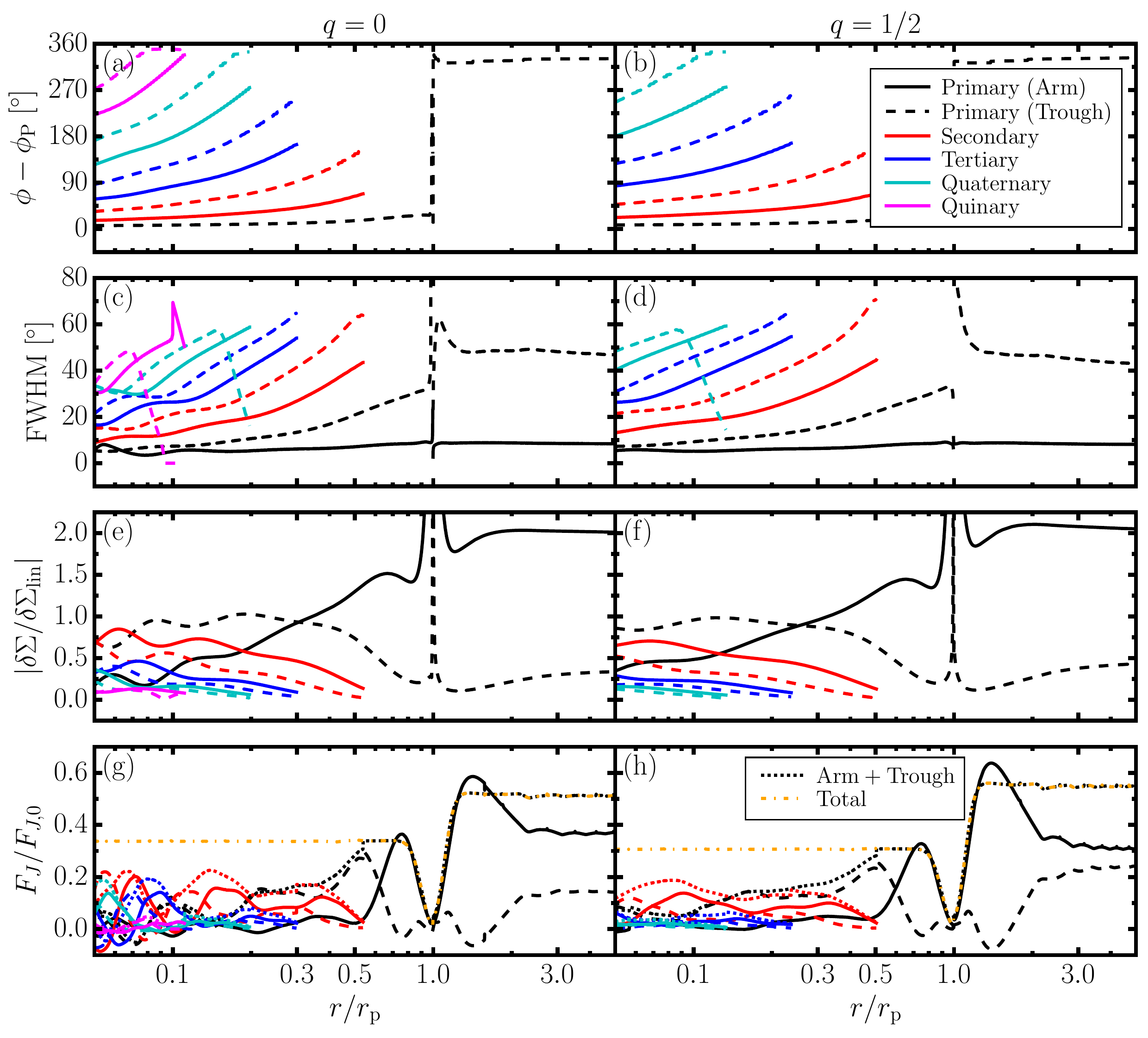}
\caption{The same as Figure~\ref{fig:arm_profiles_h}, but for different values of the temperature power law index $q$, with $h_\mathrm{p} = 0.1$.}
\label{fig:arm_profiles_q}
\end{center}
\end{figure*}

\begin{figure*}
\begin{center}
\includegraphics[width=0.99\textwidth,clip]{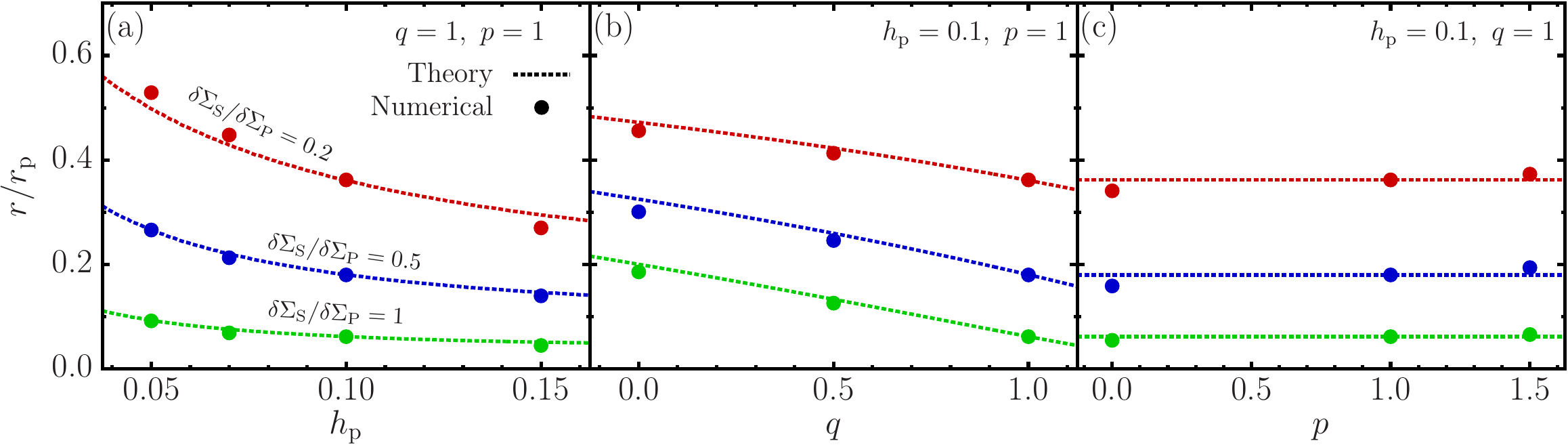}
\caption{The radii at which the amplitude of the surface density perturbation associated with the secondary spiral arm, $\delta\Sigma_\mathrm{S}$, relative to the amplitude of the primary arm, $\delta\Sigma_\mathrm{P}$, is equal to $0.2$ (red points), $0.5$ (blue points), and $1$ (green points). In each panel, two of three disk parameters (aspect ratio $h_\mathrm{p}$, temperature power law index $q$, and surface density power law index $p$) are fixed at their fiducial values, while the value of the third parameter is varied. The dashed curves indicate the theoretical predictions given by equations~(\ref{eq:dphi_criterion})--(\ref{eq:arm_width_theory}), calibrated to the fiducial disk parameters (i.e., using $h_\mathrm{p} = 0.1$).}
\label{fig:secondary_amplitude}
\end{center}
\end{figure*}

We now characterize the morphological properties---amplitude, width, and arm-to-arm separation---of the spiral arms, the radial dependence of these properties, and how they vary with the disk parameters. We first briefly describe the procedure for identifying and characterizing the spiral arms, the results of which are illustrated in Figures~\ref{fig:arm_profiles_h}--\ref{fig:arm_profiles_q}. We emphasize that the procedure for decomposing $\delta\Sigma$ into discrete spiral arms and troughs is heuristic and not unique; in practice this decomposition can be done using different criteria than the ones we used. Therefore the results should be taken as semi-quantitative descriptions.

\subsubsection{Identification and Characterization of Spiral Arms}
\label{sect:arms_method}

At $r = r_\mathrm{p}$, $\delta\Sigma$ has a single (global) maximum, located at $\approx \phi_\mathrm{p}$, which we identify as the primary arm. By following the position of this maximum across different radii, we obtain the position of the primary arm as a function of $r$. If at some $r$, a second (local) maximum with an amplitude equal to at least $10\%$ of the amplitude of the primary arm is present, it is identified as the secondary spiral arm. This relative amplitude threshold is necessary in order to avoid spurious ``detections'' of additional spirals. The position of the secondary is then followed across different radii in the same manner as the primary. The emergence of subsequent (e.g., tertiary, quaternary) arms is determined by a similar criterion, except we require the amplitude to be at least $10\%$ of the {\it strongest} spiral arm at a given $r$, which may not necessarily be the primary (since it amplitude decays with distance from the planet). The positions of the spiral arms are denoted $\phi_\mathrm{P}$, $\phi_\mathrm{S}$, and so on.

We also identify the troughs, i.e., local minima of $\delta\Sigma$. We associate each trough with a spiral arm at each $r$, since features tend to appear in arm/trough pairs (see Figure~\ref{fig:wake_profiles}). The troughs are therefore designated as the ``primary trough'', ``secondary trough'', and so on. Note that near the planet, the primary arm is accompanied by two troughs, one leading and one trailing. Toward the outer disk, the trailing trough becomes the stronger of the two, and so it is identified as the primary trough. In the inner disk, the leading trough becomes stronger, and so it is identified as the primary trough instead. This is the origin of the discontinuity in the position of the primary trough in Figs.~\ref{fig:arm_profiles_h}a--c and \ref{fig:arm_profiles_q}a--b. Multiple arms formed in the inner disk are each similarly accompanied by a leading trough. In the outer disk, when multiple arms are present (we find at most two, and only for our thinnest disk), they are instead accompanied by trailing troughs. 

Using the profiles of $\delta\Sigma$ in the vicinity of the peaks and troughs, we quantify the amplitudes and widths of the arms/troughs using the maximum and the full width at half maximum (FWHM) of $\delta\Sigma$ for arms, and of $-\delta\Sigma$ for troughs. The arm amplitudes are denoted by $\delta\Sigma_\mathrm{P}, \delta\Sigma_\mathrm{S}$, etc.

Note that since troughs are simply identified as local minima, it is sometimes the case (for high order troughs, e.g., tertiary and beyond) that $\delta\Sigma > 0$ at the trough location. In this case the width of the trough is undefined, since the amplitude is measured relative to zero. But as the trough evolves, eventually $\delta\Sigma$ at the minimum becomes negative, so the trough has a well-defined (but narrow) width. The trough width increases as its amplitude grows, until it resembles the initial widths of lower-order troughs, and then becomes more narrow towards the inner disk, in accordance with the behavior of the other arms/troughs. This explains the anomalous behavior of some of the high-order trough widths seen in, e.g., Figure~\ref{fig:arm_profiles_h}f and \ref{fig:arm_profiles_q}c--d.

\begin{figure*}
\begin{center}
\includegraphics[width=0.99\textwidth,clip]{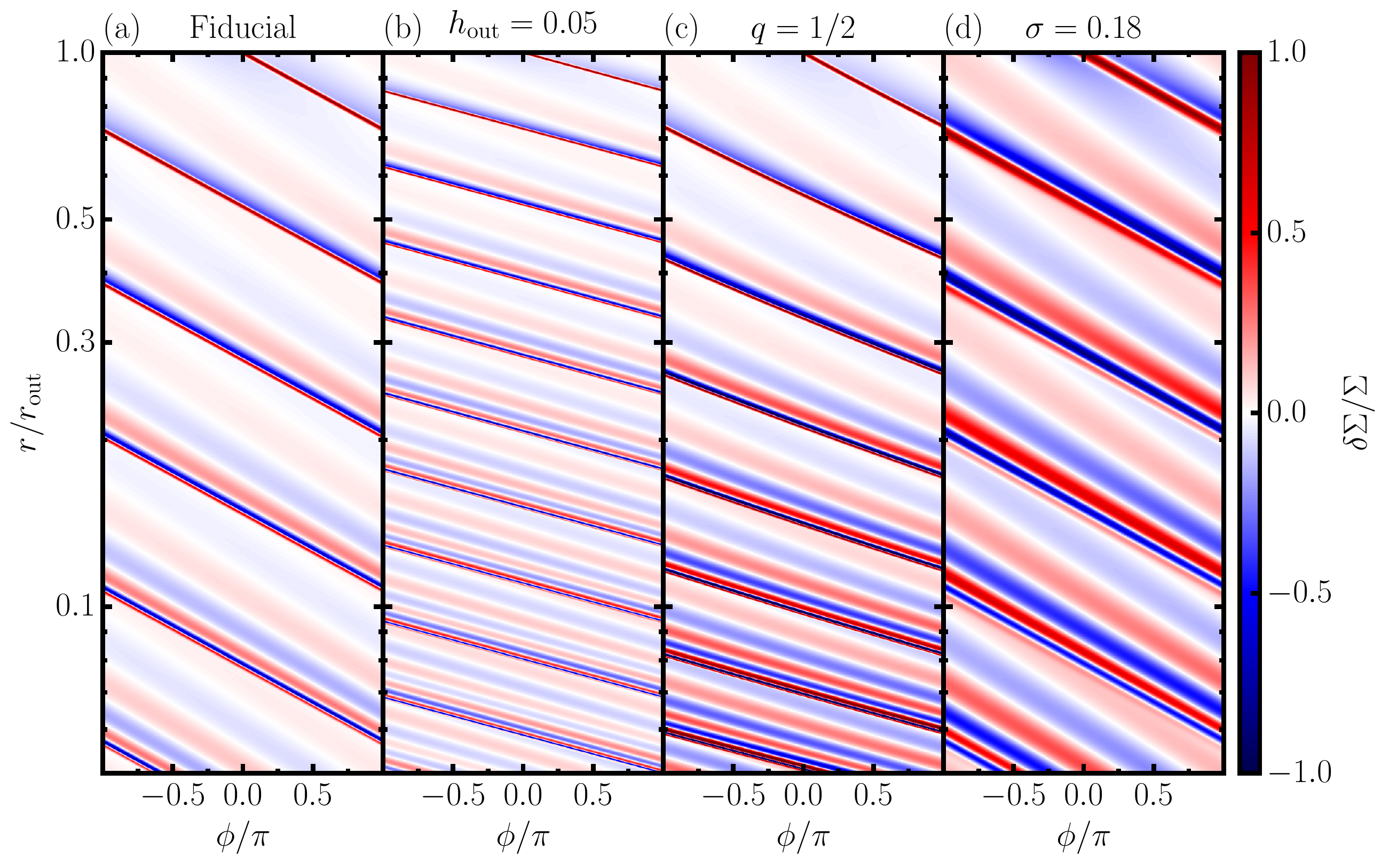}
\caption{Two-dimensional fractional surface density perturbations for the propagation of passive spiral wake with an initially specified profile at $r_\mathrm{out}$ (see \S \ref{sec:passive}). The leftmost panel shows the case of the fiducial disk parameters $h_\mathrm{out} = 0.1$, $q = 1$, and $p = 1$, and $\sigma = 0.06$. In the other panels, the disk aspect ratio, temperature power law index, and initial width are varied. In all cases, the single spiral arm imposed at $r_\mathrm{out}$ splits up into two or more arms as it propagates inwards.}
\label{fig:sigma2d_homogeneous}
\end{center}
\end{figure*}

We also keep track the AMF $F_J$ carried by each arm or trough by integrating the AMF distribution (equation~\ref{eq:dFJdphi}) over its azimuthal extent, which is delineated by the zeros of $\delta\Sigma$. These roughly correspond to the zeros of $f_J$, since the latter is approximately proportional to $(\delta\Sigma)^2$ (see Fig.~\ref{fig:wake_profiles} and equation~\ref{eq:dFJdphi_approx}). Note that the AMF is divided up and assigned to the arms and troughs in a conservative manner, so that the sum of the AMF associated with all of the peaks and troughs at a given radius is equal to $F_J(r)$.

The extracted properties of the arms and troughs (position, width, amplitude, and AMF) are shown for the fiducial case and for cases with different disk aspect ratios in Figure~\ref{fig:arm_profiles_h}, and for cases with different temperature power law indices in Figure~\ref{fig:arm_profiles_q}. Note that the arm/trough positions are shown relative to the position of the primary arm, $\phi_\mathrm{P}$, which is well-approximated by equation~(\ref{eq:phi_lin}) (although it may differ by $\sim 10^\circ$ far from the planet, see Fig.~\ref{fig:wake_profiles}). Also note that the amplitudes are scaled by $\delta\Sigma_\mathrm{lin}$, in order to remove the simple variation of amplitude with $r$ resulting from angular momentum flux conservation. The AMF (lower panels) is scaled by the characteristic AMF due to the one-sided Lindblad torques given by Equation (\ref{eq:FJ0}).

\subsubsection{Outer Disk}
\label{sect:arms_outer}

The structure of the surface density perturbation (see Figs.~\ref{fig:arm_profiles_h}g--i and \ref{fig:arm_profiles_q}e--f) in the outer disk is fairly simple. Typically the perturbation consists only of a primary arm, with a width of about $10^\circ$, and its associated trough, which is about four times wider. Their amplitudes vary slowly, with the gradual decay of the arm amplitude accompanied by the gradual growth of the trough amplitude. The amplitude of the trough remains small relative to the arm (a few tenths at most), although it carries a significant fraction of the AMF (see Figs.~\ref{fig:arm_profiles_h}j--l and \ref{fig:arm_profiles_q}g--h) due to its large width. For $h_\mathrm{p} = 0.1$ and smaller (with $q = 1$), the trough carries most of the AMF (i.e., more than the peak) for $r \gtrsim (1.5 - 2.0) r_\mathrm{p}$. 

For the thinnest disk we have considered ($h_\mathrm{p} = 0.05$), a secondary arm does form at about $3 r_\mathrm{p}$ (see Fig.~\ref{fig:arm_profiles_h}g), which is explained in \S \ref{subsec:theory_inner_outer}. Its amplitude grows very slowly with $r$, and is only about $12\%$ of that of the primary at $5 r_\mathrm{p}$. Note that we have solved for the structure of the perturbations only out to $5 r_\mathrm{p}$, and so it is possible that a secondary arm emerges beyond this radius for larger $h_\mathrm{p}$ as well (although our analytic estimates indicate that this is unlikely, see \S \ref{subsec:theory_inner_outer}). If so, it forms very far from the planet in comparison to the inner disk, where a secondary arm always forms at about half of the orbital radius of the planet. And as demonstrated by the $h_\mathrm{p} = 0.05$ case, the outer secondary spiral is very weak even when it exists. Therefore we do not devote much attention to secondary spirals in the outer disk.

\subsubsection{Inner Disk}
\label{sect:arms_inner}

In the inner disk, multiple (three to five) spiral arms are robustly formed for a variety of parameters. For the fiducial parameters (Figure~\ref{fig:arm_profiles_h}b,e,h,k), the secondary arm first appears at $0.46 r_\mathrm{p}$, with an initial separation of about $70^\circ$ from the primary arm, and an initial width of $45^\circ$. At smaller radii, its separation relative to the primary arm decreases to about $30^\circ$, and its width decreases to about $20^\circ$. The amplitude of the secondary arm increases as that of the primary decreases. At $r = 0.17 r_\mathrm{p}$, the amplitude of the secondary is about half of the primary, and at $r = 0.06 r_\mathrm{p}$, the secondary amplitude exceeds the primary amplitude. 

A weak tertiary arm forms at about $0.3 r_\mathrm{p}$, and a weak quaternary arm forms at about $0.1 r_\mathrm{p}$. The tertiary and quaternary arms each form at $\approx 135^\circ$ from the previous arm (roughly twice the secondary-to-primary separation), with an initial width of about $60^\circ$, see Figure \ref{fig:arm_profiles_h}b,e. After the quaternary arm forms, roughly the entire $2\pi$ extent of the disk is populated with spirals, and so it is unlikely that any more well-defined arms would form\footnote{Although since the arms become more narrow toward the inner disk, there may possibly be room for additional arms at even smaller radii.}. Nonetheless, we emphasize that the tertiary and higher-order arms have very small amplitudes relative to the primary and secondary even at $r = 0.05 r_\mathrm{p}$ (see Figure \ref{fig:arm_profiles_h}h), and thus are relatively unimportant features.

As in the outer disk, the primary trough carries a significant portion of the total AMF (see Figs.~\ref{fig:arm_profiles_h}j--l and \ref{fig:arm_profiles_q}g--h), and typically carries more AMF than the primary arm at $r\lesssim (0.5 - 0.8) r_\mathrm{p}$, which is a larger radius than the one at which the secondary arm forms, or at which the secondary arm amplitude becomes stronger than the primary (in terms of amplitude). The AMF of the secondary arm exceeds primary arm AMF for $r\lesssim (0.4 - 0.5) r_\mathrm{p}$. Far inside inner disk ($r \approx 0.1 r_\mathrm{p}$), the AMF is primarily carried by secondary arm, secondary trough, and primary trough, while contributions from the other peaks and troughs are negligible.


\subsection{Dependence on Disk Parameters}
\label{sect:par_dep}


\begin{figure}
\begin{center}
\includegraphics[width=0.49\textwidth,clip]{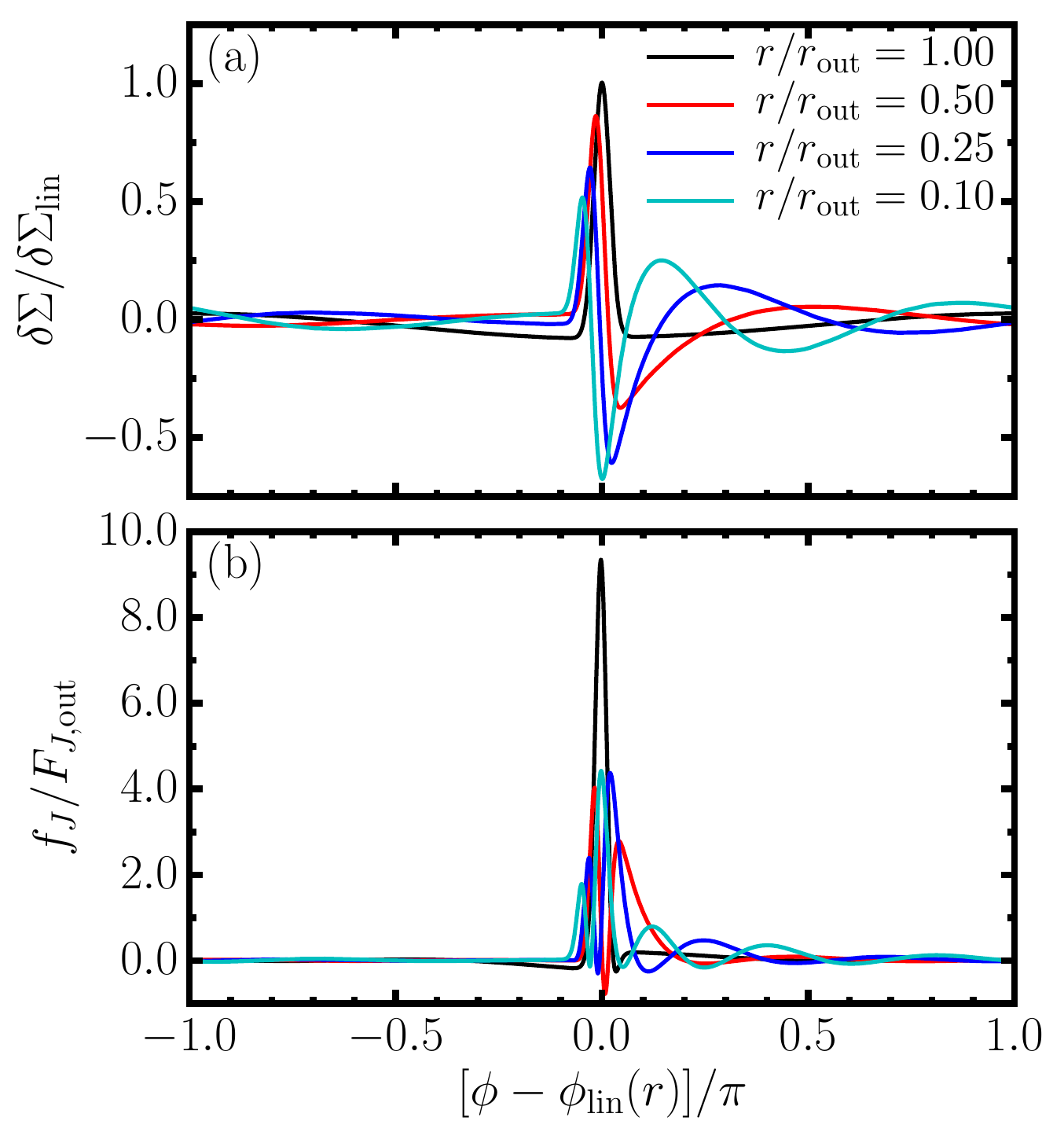}
\caption{Profiles of the surface density perturbation (top panel) and azimuthal distribution of angular momentum flux (bottom panel) at different radii, as in Figure~\ref{fig:wake_profiles}, but for the homogeneous problem of a passive spiral wake.}
\label{fig:wake_profiles_homogeneous}
\end{center}
\end{figure}

\begin{figure*}
\begin{center}
\includegraphics[width=0.99\textwidth,clip]{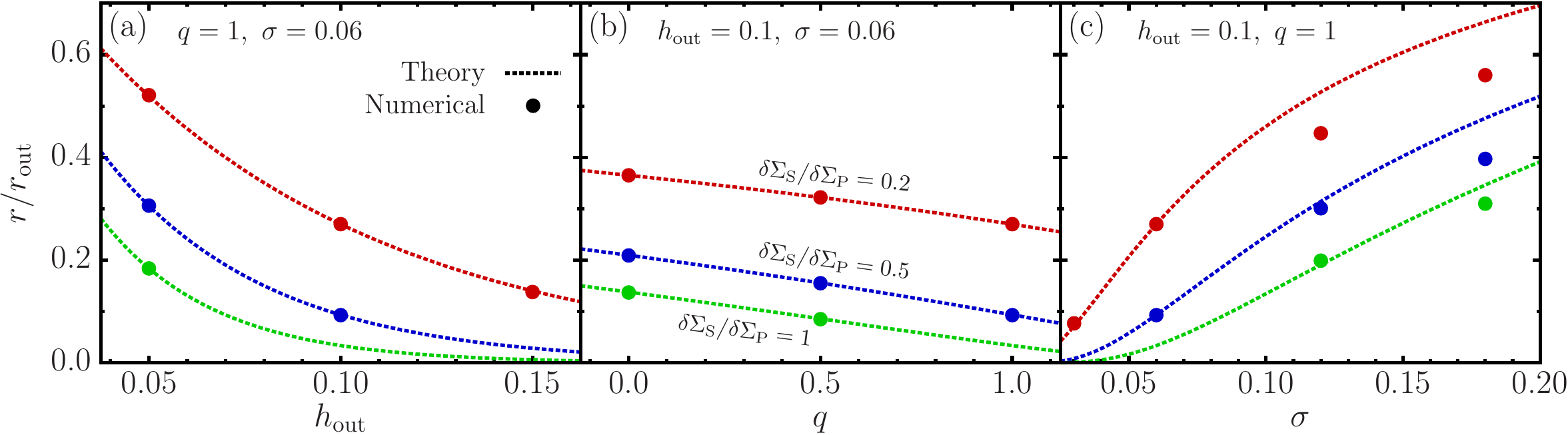}
\caption{The radii at which the amplitude of the secondary arm reaches $0.2$ (red points), $0.5$ (blue points), and $1$ (green points) relative to that of the primary arm, as in Figure~\ref{fig:secondary_amplitude}, but for the homogeneous problem. In the different panels, the aspect ratio, temperature power law index, and initial perturbation width are varied while the other parameter values are held fixed. Note that ``missing'' points indicate that the secondary amplitude did not reach the specified threshold anywhere for $r > r_\mathrm{in} = 0.05$ for the specified parameter value. The dashed curves show the theoretical predictions given by equations~(\ref{eq:dphi_criterion}) and (\ref{eq:dphin_passive}), calibrated to the fiducial parameters.}
\label{fig:secondary_amplitude_homogeneous}
\end{center}
\end{figure*}

Figure \ref{fig:sigma2d}a,c shows that lowering $h_p$ results in more tightly wound spiral arms. For such colder and thinner disks, the secondary arm (in the inner disk) forms closer to the primary (in azimuthal angle), with a smaller initial width and smaller initial arm-to-arm separation, and it becomes stronger than the primary also at a larger radius, see Figure~\ref{fig:arm_profiles_h}a,d,g,j. 

For temperature profiles with $q$ different from unity, Figure \ref{fig:sigma2d}a,b shows that the pitch angle of the spirals vary with radius. For $q < 1$, the initial properties of the secondary arm are relatively unchanged, but its amplitude becomes larger than the primary amplitude at larger radii (similar to the case of lower $h_p$), see Figure~\ref{fig:arm_profiles_q}. Additional (tertiary and beyond) arms also form closer to the planet for thinner disks, and more arms are formed---as many as five for $h_\mathrm{p} = 0.05$ (however, the full azimuthal extent of the inner disk is not fully populated with spirals in this case, as it is for larger aspect ratios).

The development of the secondary arm in the inner disk is summarized in Figure~\ref{fig:secondary_amplitude} for different disk parameters. It shows the radial location at which the amplitude of the secondary arm, $\delta\Sigma_\mathrm{S}$, relative to the amplitude of the primary, $\delta\Sigma_\mathrm{P}$, is equal to $0.2$, $0.5$, and $1$. In each panel, two of the three disk parameters ($h_\mathrm{p}$, $q$, and $p$) are held fixed while the third one is varied. These properties depend most strongly on $h_\mathrm{p}$, with the arm forming closer to planet for smaller $h_\mathrm{p}$. For values of $q$ smaller than the fiducial $q = 1$, i.e., for more flared disks, the secondary arm forms somewhat closer to the planet. However, its amplitude grows with radius faster than the case with $q = 1$, reaching $50\%$ and $100\%$ of the primary amplitude at significantly larger radii.

The development of the secondary spiral is relatively insensitive to the slope of the surface density profile, for a plausible range of values ($0 < p < 3/2$). This is evident in the structure of the 2D surface density maps in Figures~\ref{fig:sigma2d}--\ref{fig:sigma2d_cartesian}, which show the cases $p = 1$ and $p = 0$ (note that the overall scaling of the amplitude of the perturbations with $r$ differs between the two cases, but this is trivially described by AMF conservation, see equation~\ref{eq:amp_lin}). This is shown quantitatively in Figure~\ref{fig:secondary_amplitude}c. The radius at which the secondary spiral forms differs by only about $5\%$ between the cases $p = 0$ and $p = 3/2$. 


\section{Results for Passively Propagating Spiral Waves}
\label{sec:passive}


As will be shown in \S \ref{sec:analytic}, the formation of the secondary spiral can be described fairly well by the local (WKB) approximation, which is valid far from resonances, where multiple spiral arms emerge, and where the gravitational influence of the planet is negligible. This indicates that the external potential of the planet plays a minimal role in this process. Therefore, motivated by the results of \citet{AR18}, we may expect that any spiral wake formed by the phase coherence of a series of modes launched in a companion-free disk will also break up into multiple spirals.

To this end, we investigate the passive linear propagation of a spiral wake (not subject to the external potential of a planet). More specifically, we perform a calculation representing a \textit{linear analogue} of the nonlinear calculation of \citet{AR18}, closely following their setup. In this setup, a surface density perturbation $\delta\Sigma(\phi)$ rotating with a pattern frequency $\omega_\mathrm{p}$ is imposed at a radius $r_\mathrm{out}$, and is allowed to propagate towards the disk center (only inward propagation is allowed provided that $\omega_\mathrm{p} < \Omega_\mathrm{out}$).

We solve for the structure of the modes of the homogeneous wave equation (\ref{eq:master_eqn}) with $\Phi_m = 0$ by shooting solutions inward from $r_\mathrm{out}$ in order to satisfy the outgoing wave boundary condition at $r_\mathrm{in} = 0.05 r_\mathrm{out}$. For simplicity, here we choose a mode pattern frequency $\omega_\mathrm{p} = 0$ (so that the wave pattern is fixed in the inertial frame), which corresponds to waves launched at a radius $\gg r_\mathrm{out}$. Once the mode profiles are obtained, any azimuthal profile $\delta\Sigma$ can be constructed from an appropriate linear combination of the different $\delta\Sigma_m$'s.

The perturbation imposed at $r_\mathrm{out}$ is chosen to be a Gaussian\footnote{The actual profile imposed at $r_\mathrm{out}$ differs slightly from the form specified by equation~(\ref{eq:pert_out}). Specifically, the actual profile is ``missing'' the $m = 1$ Fourier component. This is because the $m = 1$ mode is evanescent interior to its OLR (which is assumed to be located at $r \gg r_\mathrm{out}$), and exterior to its ILR (which is formally located at $r = 0$). Therefore, the amplitude of the $m = 1$ mode is assumed to be zero at $r_\mathrm{out}$, as a result of its evanescent propagation from large $r$. Note that if we had instead chosen the amplitude of the $m = 1$ mode to be finite at $r_\mathrm{out}$ (so that $\delta\Sigma$ is given exactly by eq.~\ref{eq:pert_out}), it would decay away towards small $r$ anyway, and it would not strongly affect our results.},
\be
\label{eq:pert_out}
\delta\Sigma(r_\mathrm{out}) \propto \exp\left[-\frac{1}{2}\left(\frac{\phi}{\sigma}\right)^2\right].
\ee
Thus, there are four parameters in our setup. These are the three disk parameters $h_\mathrm{out}$ and $q$ which define the sound speed profile, now parameterized by $c_\mathrm{s}(r) = h_\mathrm{out} r_\mathrm{out}\Omega_\mathrm{out}(r/r_\mathrm{out})^{-q/2}$, and the surface density power law index $p$, as well as the initial azimuthal width $\sigma$ of the surface density perturbation. For the fiducial parameters, we choose $h_\mathrm{out} = 0.1$, $q = 1$, $p = 1$, and $\sigma = 0.06$. The fiducial $\sigma$ is chosen to produce a surface density profile similar to the one produced near the planet for the case of the fiducial parameters in the inhomogeneous problem.

Figure~\ref{fig:sigma2d_homogeneous} shows 2D maps of the fractional surface density perturbation, for the fiducial parameters, and for several cases with varied parameters. In all of these cases, the initial, single peaked perturbation profile breaks up into multiple arms, in agreement with findings of \citet{AR18}. The azimuthal profiles of $\delta\Sigma$ at different radii are shown in Figure~\ref{fig:wake_profiles_homogeneous} for the fiducial parameters. They are qualitatively very similar to the profiles that develop in the inner disk in the case of the spiral wake driven by a planet, see Figure~\ref{fig:wake_profiles}. The azimuthal distribution of the angular momentum flux similarly becomes divided up into several different peaks. We emphasize that the appearance of multiple spirals in this setup, in which there are no resonances in the disk, indicates that the formation of multiple spirals is not connected to resonances.

Figure~\ref{fig:secondary_amplitude_homogeneous} summarizes the dependence of secondary arm amplitude on disk parameters and the initial width of the spiral wake. The dependence on the surface density profile index $p$ is extremely weak. In fact, it is much weaker than in the case with a planet: the radius at which the secondary spiral emerges changes by less than $1\%$ when the value of $p$ is varied between $0$ and $1$. This lack of sensitivity to $\Sigma$ profile is explained in \S \ref{sec:analytic}.

On the other hand, properties of the secondary spiral depend strongly on the initial width of the perturbation, see Figure \ref{fig:secondary_amplitude_homogeneous}c. Wider initial perturbations lead to the development of the secondary spiral at larger radii, which is discussed in \S \ref{sec:analytic}. In this experiment, we have isolated the effect of the initial perturbation width, as it is controlled by the parameter $\sigma$, which is unrelated to the disk parameters. In the case of a density wave launched by a planet, the width of the spiral wake near the planet is set by the disk aspect ratio, since the perturbation is dominated by modes with $m \sim 1/h_\mathrm{p}$, and so thinner disks produce narrower initial wakes. In the homogeneous case, by separating these two effects, we see that variation of the disk aspect ratio has an even stronger effect on the emergence of the secondary spiral than it does in the case with a planet: changing $h_\mathrm{out}$ by a factor of three changes the radius at which the secondary emerges by a factor of two. Evidently these two effects somewhat cancel out in the case with a planet.


\section{Analytical understanding of our results}
\label{sec:analytic}


In this section, we present theoretical arguments, following those made by \citet{OL02} and \citet{BZ18a}, which interpret spiral arm formation as resulting from constructive interference among different azimuthal modes. We then use these arguments to explain and interpret the results of our numerical calculations.


\subsection{Mode Phases and Interference}
\label{subsec:phase_bz}


The surface density perturbation of the mode with azimuthal number $m$ is
\be
\delta\Sigma_m(r,\phi) = |\delta\Sigma_m(r)| \exp[\mathrm{i}\psi_m(r)],
\ee
where
\be
\psi_m(r) = \mathrm{Arg}[\delta\Sigma_m(r)] + m(\phi-\phi_\mathrm{p}).
\ee
The phase of $\delta\Sigma_m$ can be expressed as \citep{OL02} 
\be
\label{eq:exact_phase}
\mathrm{Arg}[\delta\Sigma_m(r)] = \mathrm{sgn}(r-r_\mathrm{p})\frac{\pi}{4} + \int_{r_\pm}^r k_m(r^\prime)\mathrm{d}r^\prime,
\ee
where 
\be
r_\pm = \left(1\pm\frac{1}{m}\right)^{2/3} r_\mathrm{p}
\ee
is the location\footnote{Note that this expression neglects $\mathcal{O}(h_\mathrm{p}^2)$ terms due to radial pressure support and non-zero $N_r$.} of the outer ($+$)/inner ($-$) Lindblad resonance, and the $\pi/4$ term is the phase shift associated with the resonance. The radial number is given in the local (WKB) limit (and in the Keplerian limit $\kappa = \Omega$) by
\be
\label{eq:km}
k_m(r) = \frac{1}{c_\mathrm{s}} \left[m^2(\Omega-\Omega_\mathrm{p})^2 - \Omega^2\right]^{1/2}.
\ee
Note that for $m \gg 1$, $k_m(r) \approx \tilde{k}_m(r)$, where
\be
\tilde{k}_m(r) = \frac{m}{c_\mathrm{s}}|\Omega-\Omega_\mathrm{p}|,
\ee
which is simply proportional to $m$. Defining
\be
\Delta k_m(r) = k_m(r) - \tilde{k}_m(r),
\ee
we can write
\be
\label{eq:psi_m}
\begin{aligned}
\psi_m & = \mathrm{sgn}(r-r_\mathrm{p})\frac{\pi}{4} + m(\phi-\phi_\mathrm{P}) \\
& + \int_{r_\pm}^r \Delta k_m(r^\prime) \mathrm{d}r^\prime - \int_{r_\mathrm{p}}^{r_\pm} \tilde{k}_m(r^\prime) \mathrm{d}r^\prime.
\end{aligned}
\ee
Here we have defined\footnote{Note that $\phi_\mathrm{P}$ is equivalent to $\phi_\mathrm{lin}$ as defined in equation~(\ref{eq:phi_lin}). Here it has been redefined to emphasize that it gives a theoretical prediction for the azimuthal position of the primary spiral arm.}
\ba
\phi_\mathrm{P} &=&  \phi_\mathrm{p} - \int_{r_\mathrm{p}}^r\tilde{k}_m(r^\prime) \mathrm{d}r^\prime
\nonumber\\
&=&\phi_\mathrm{p} - \int_{r_\mathrm{p}}^r \frac{|\Omega(r^\prime)-\Omega_\mathrm{p}|}{c_\mathrm{s}(r^\prime)} \mathrm{d}r^\prime.
\label{eq:primary_pos}
\ea  
Note the distinction between $\phi_\mathrm{p}$ (with a lowercase subscript), the position of the planet, and $\phi_\mathrm{P}$ (with an uppercase subscript), the position of the primary spiral arm. Also, in equations (\ref{eq:exact_phase}), (\ref{eq:psi_m}), (\ref{eq:primary_pos}) note the different limits of integration. 

The perturbation $\delta\Sigma_m$ has a maximum found by setting $\psi_m = 0$. For modes with $m \gg 1$, the last two terms in equation~(\ref{eq:psi_m}) become negligible, so that the maxima of these modes have positions $\phi \approx \phi_\mathrm{P}$. For general $m$, the maximum of $\delta\Sigma_m$ interferes constructively with these modes if
\be
\int_{r_\pm}^r \Delta k_m(r^\prime) \mathrm{d}r^\prime - \int_{r_\mathrm{p}}^{r_\pm} \tilde{k}_m(r^\prime) < \Delta\phi_0,
\label{eq:clust}
\ee
where $\Delta\phi_0$ is the maximum phase difference which results in constructive interference. In general, this criterion is indeed satisfied (for some range of $r$ and values of $m$), resulting in the formation of the primary spiral arm. Therefore, $\phi_\mathrm{P}$ gives the approximate position of the primary spiral \citep{OL02,R02}. 
To understand the origin of higher-order arms (secondary, tertiary, etc.), let us note that maxima of  $\delta\Sigma_m$ are also attained at $\psi_m = 2\pi n$, where $n = 1, 2, \ldots, m - 2, m - 1$ is an integer. Maxima in this range of $n$ are distinct in a sense that, at a given $r$, each of them corresponds to a well defined azimuthal location $\phi = \phi_{m,n}$ (in a frame co-rotating with the planet), where
\be
\label{eq:phimn_defn}
\phi_{m,n}(r) = \phi_\mathrm{p} + \frac{1}{m}\left[-\mathrm{Arg}(\delta\Sigma_m(r)) + 2\pi n\right].
\ee
Values of $n$ outside the interval $[1,m-1]$ yield $\phi_{m,n}$ coinciding with one of the azimuthal locations inside this interval.

As pointed out by \citet{BZ18a}, these locations may define additional curves along which constructive interference occurs. In their formulation, a spiral arm may be formed as a result of constructive interference among the $\phi_{m,n}$ peaks with different values of $m$ (but a fixed value of $n$). The primary spiral is a result of constructive interference of peaks with $n = 0$. In the inner disk, the secondary spiral is a result of interference of peaks with $n = 1$, the tertiary spiral is associated with $n = 2$, and so on. In the outer disk, $n = m - 1$ peaks are associated with the formation of a secondary spiral, $n = m - 2$ with a tertiary spiral, and so on.

\begin{figure*}
\begin{center}
\includegraphics[width=0.99\textwidth,clip]{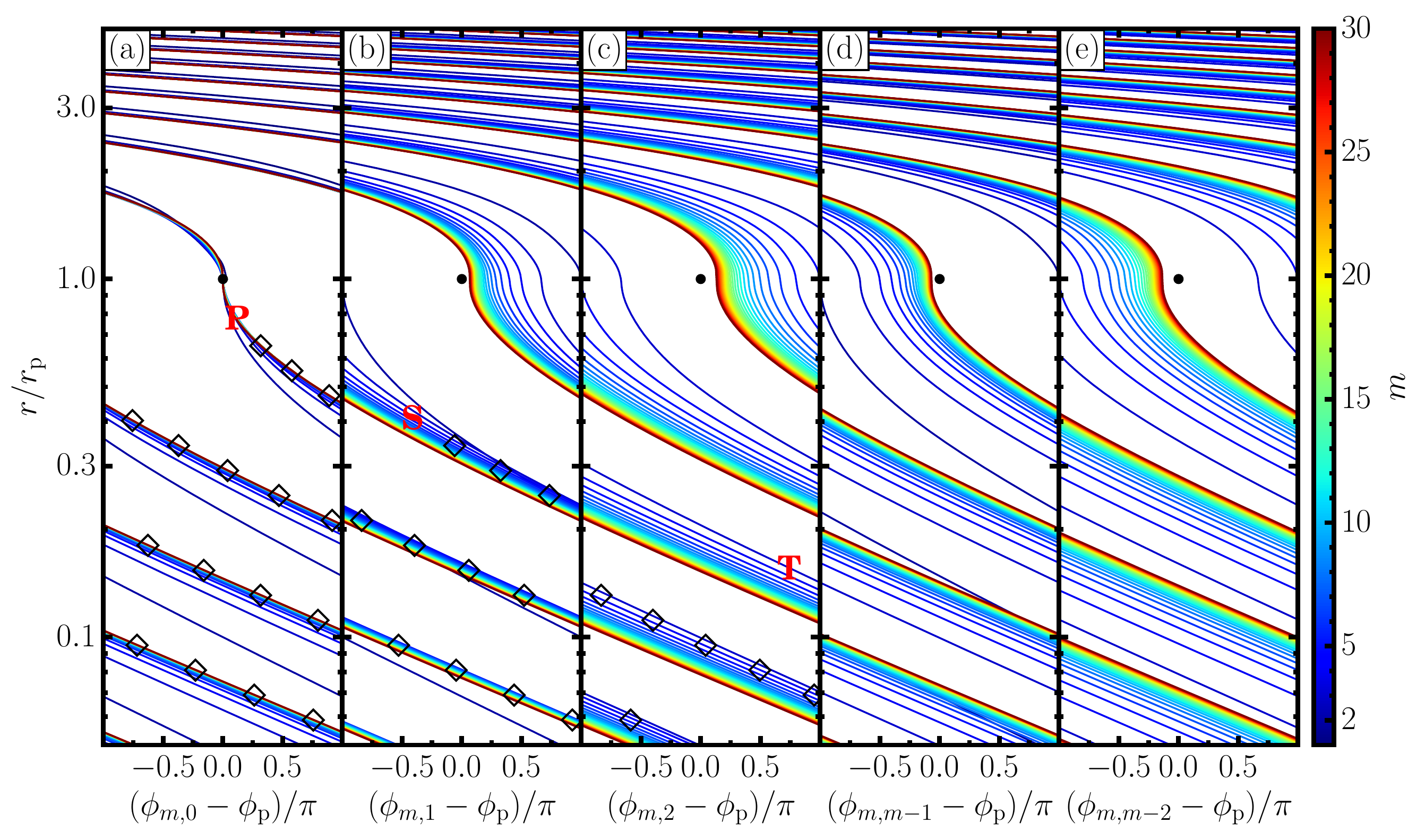}
\caption{The phases of different surface density peaks associated with modes with different azimuthal numbers. Each $\phi_{m,n}$ corresponds to the $n$th peak associated with the mode of azimuthal number $m$. Constructive interference between peaks with different values of $m$ but the same value of $n$ may be responsible for the formation of the multiple spiral arms. The different colored curves correspond to different azimuthal numbers, and the different panels correspond to different values of $n$. In the three leftmost panels, the phases of the $n = 0$, $n = 1$, and $n = 2$ peaks, which are associated the primary, secondary, and tertiary arms (respectively), are shown, along with symbols (diamonds) indicating the actual location of these arms. In the two rightmost panels, phases of peaks with $n = m - 1$ and $n = m - 2$ are shown. These peaks are not relevant to secondary spiral arm formation in the inner disk, as they are located behind, rather than in front of, the primary arm.}
\label{fig:mode_phases}
\end{center}
\end{figure*}

The argument of \citet{BZ18a} is based on the WKB approximation for the phases of the modes, and no account is given to the behavior of the mode amplitudes. In our calculations, the mode phases (as well as amplitudes) are computed exactly, by solving for the {\it global} mode structure (i.e., not using WKB approximation). To illustrate how the mode interference idea works in our fully self-consistent calculation, in Figure~\ref{fig:mode_phases} we show the phases of different crests of the modes defined by equation~(\ref{eq:phimn_defn}), which were numerically computed using our full linear solutions. In agreement with \citet{OL02}, this figure indicates that the mode phases with $n = 0$ are tightly clustered in phase, and these phases closely follow that of the primary arm. But in addition to that, one also sees that the phases with $n = 1$, which are initially very spread out (over a range of $\sim \pi$), become more clustered towards the inner disk. The phases of these modes follow the phase of the secondary spiral arm, indicating that the constructive interference of $n = 1$ mode crests is indeed responsible for the emergence of the secondary arm. 

The $n = 2$ phases also start out very spread out in azimuth at $r = r_\mathrm{p}$. They slowly converge as $r$ decreases, although they still span a range of $\sim \pi/2$ even at $r = 0.05 r_\mathrm{p}$. These phases are approximately coincident with that of the tertiary arm, although this correspondence is not as tight as it is for the cases of the primary and secondary arms. This highlights the fact that the phase information of the modes, while suggestive of the structure of the spiral arms, is not sufficient to fully capture their structure. Rather, a {\it full consideration of the mode phases and amplitudes is required}, which is done in this work.

None of the phases {\it behind} the $n = 0$ phases (i.e. $n = m - 1, n = m - 2$) are as tightly clustered as those with $n = 0$. Correspondingly, no spiral arm forms behind the primary in the inner disk (for the fiducial parameters). 

\subsection{Analytic Estimates}


We now try to predict the radius at which secondary arm emerges, as well as its location and width, using a phase interference argument. We do this by estimating the spread in the relevant $\phi_{m,n}$'s, and identifying the regions in which they are appropriately clustered as the locations of the different spiral arms. In order to do this, we make use of the WKB approximation, which we find to reproduce the mode phases ($\phi_{m,n}$'s) found in our numerical calculations with reasonable accuracy. For modes with $m$ close to $m_* \approx (2h_\mathrm{p})^{-1}$, the error in the WKB phases is $\sim 20\%$ in the vicinity of the Lindblad resonances, and an order of magnitude smaller far from the resonances, where secondary spiral arm formation occurs.

From (\ref{eq:exact_phase}) and (\ref{eq:phimn_defn}), we have
\be
\phi_{m,n} = \phi_\mathrm{p} - \mathrm{sgn}(r-r_\mathrm{p})\frac{\pi}{4m} + \frac{2\pi n}{m} - \int_{r_\pm}^r \frac{k_m(r^\prime)}{m}\mathrm{d}r^\prime
\label{eq:phimn}
\ee
(cf. \citealt{BZ18a}). The planet-induced density wake is dominated by modes with $m \approx m_*$. Therefore, defining $\phi_n \equiv \phi_{m_*,n}$, we follow the ansatz of \citet{BZ18a}, and identify in the inner disk the position of the secondary spiral as $\phi_\mathrm{S} = \phi_1$, the position of the tertiary spiral as $\phi_\mathrm{T} = \phi_2$, and so on. The positions of the secondary and higher-order spiral arms relative to the primary arm are then given approximately by
\be
\begin{aligned}
\label{eq:arm_sep_theory}
\phi_n - \phi_\mathrm{P} & = -\mathrm{sign}(r-r_\mathrm{p})\frac{\pi}{4m_*} + \frac{2\pi n}{m_*} \\
& - \int_{r_\pm}^r \frac{\Delta k_{m_*}(r^\prime)}{m_*}\mathrm{d}r^\prime + \int_{r_\mathrm{p}}^{r_\pm} \frac{\tilde{k}_{m_*}(r^\prime)}{m_*}\mathrm{d}r^\prime,
\end{aligned}
\ee
see equations (\ref{eq:primary_pos}) and (\ref{eq:phimn}).

However, in order for a spiral arm to exist at $\phi = \phi_n$, the appropriate $\phi_{m,n}$'s must be sufficiently clustered. The spread in $\phi_{m,n}$ for values of $m$ in the vicinity of $m_*$ is
\be
\label{eq:dphimn}
\delta\phi_{m,n} = \left|\left(\frac{\partial\phi_{m,n}}{\partial m}\right)_{m_*}\right|\Delta m,
\ee
where $\Delta m$ represents the range of azimuthal numbers which contribute to the arm. We assume that this range is comparable to the critical $m$ itself, taking $\Delta m = \zeta m_*$, with $\zeta \sim 1$.\footnote{This is a reasonable assumption for the narrow primary and secondary spirals. However, the broader widths of the tertiary and higher order spiral arms may indicate that they are dominated by modes with a smaller range of $m$, indicating that $\zeta < 1$ may be appropriate for these spirals.} In order for the the $\phi_{m,n}$'s to constructively interfere and form a spiral arm, we require that $\delta\phi_{m,n} < \Delta\phi_0/m_*$ (where, as in equation \ref{eq:clust}, $\Delta\phi_0$ is a critical separation required for constructive interference, so that the variation of $\psi_m$ is less than $\Delta\phi_0$). Evaluating the derivative in equation~(\ref{eq:dphimn}) and simplifying, we find the condition for spiral arm formation
\be
\label{eq:dphi_criterion}
\delta\phi_n \equiv \zeta^{-1}m_* \delta\phi_{m,n} < \Delta\tilde\phi_0,
\ee
where $\Delta\tilde\phi_0=\Delta\phi_0\zeta^{-1}$ is a new ``phase spread'' constant and\footnote{In equation~(\ref{eq:arm_width_theory}), $n$ should take on the values $0, \pm 1, \pm 2,$ etc. While the definition of $\phi_{m,n}$ (equation~\ref{eq:phimn_defn}) also permits values such as $n = m - 1, m - 2, \ldots$, note that, e.g., $\phi_{m,m-1} = \phi_{m,-1}$.}
\be
\label{eq:arm_width_theory}
\delta\phi_n = \left|2\pi n - \mathrm{sign}(r-r_\mathrm{p})\frac{\pi}{4} + \int_{r_\pm}^r \frac{\Omega^2(r^\prime)\mathrm{d}r^\prime}{c_\mathrm{s}^2(r^\prime)k_{m_*}(r^\prime)}\right|.
\ee
Note that $\delta\phi_n$ represents the spread in mode phases, while $\delta\phi_n/m_*$ gives the approximate spiral arm width (provided that a spiral arm exists). The integral in equation~(\ref{eq:arm_width_theory}) is positive (negative) in the outer (inner) disk. Therefore, a necessary condition for $\delta\phi_n$ to become small enough to form a spiral arm is that $n \le 0$ ($n \ge 0$) in the outer (inner) disk. Hence, secondary and higher order spirals can only be attributed to constructive interference of $\phi_{m,n}$ with $n > 0$ in the inner disk, and only to $\phi_{m,n}$ with $n < 0$ (or equivalently, $n = m - 1, m - 2$, etc.) in the outer disk, should spiral arms be present there \citep{BZ18a}. Note that equation~(\ref{eq:arm_width_theory}) can be used to estimate the phase spread of the primary spiral arm by taking $n = 0$. From this we see that the formation of the secondary spiral is coincident with the dissolution of the primary spiral, since $\delta\phi_0$ necessarily becomes large as $\delta\phi_1$ becomes small. This explains a trend seen in Figs.~\ref{fig:arm_profiles_h}g--i and \ref{fig:arm_profiles_q}e--f: the decrease in $\delta\Sigma_\mathrm{P}$ toward the inner disk is accompanied by an increase in $\delta\Sigma_\mathrm{S}$.

Up to this point, we have left the value of $\Delta\tilde\phi_0$ arbitrary. Roughly speaking, we expect that taking $\Delta\tilde\phi_0 \approx \pi$ in equation~(\ref{eq:dphi_criterion}) should qualitatively predict the presence of a spiral arm \citep{OL02}. In practice, we can use our full numerical results to calibrate this criterion. We compute values of $\Delta\tilde\phi_0$ corresponding to different relative strengths of the secondary (and tertiary) spiral arm. We do this by calculating using equation (\ref{eq:arm_width_theory}), for example, $\delta\phi_1(r_{\mathrm{S}20})$, where $r_{\mathrm{S}20}$ denotes location at which $\delta\Sigma_\mathrm{S}/\delta\Sigma_\mathrm{P} = 20\%$, inferred from our numerical results. These values are given in Table~\ref{tab:dphi_values}.


\subsubsection{Passive Spirals}


The same formalism can also be applied to our numerical experiments for a passively propagating spirals presented in \S \ref{sec:passive}. In this setup, we assume that all modes are exactly in phase at $r_\mathrm{out}$. We also assume that the (inner) Lindblad resonances for all modes are located exterior to $r_\mathrm{out}$. Therefore, in equation~(\ref{eq:exact_phase}), we drop the $\pi/4$ term (associated with the resonance), and take $r_\mathrm{out}$ as the lower limit of the $k_m$ integral. Therefore, we have
\be
\phi_{m,n} = \frac{2\pi n}{m} - \int_{r_\mathrm{out}}^r \frac{k_m(r^\prime)}{m} \mathrm{d}r^\prime.
\ee
In the expression for the radial wavenumber (\ref{eq:km}), the orbital frequency of the planet $\Omega_\mathrm{p}$ should be replaced by the specified pattern frequency $\omega_\mathrm{p}$. Additionally, the initial width $\sigma$ of the spiral launched at $r_\mathrm{out}$ is explicitly specified rather than being set by disk aspect ratio as it was for a wake launched by a planet. Therefore, we take $m_* \approx (2\sigma)^{-1}$ as an approximation of the dominant azimuthal mode number in the homogeneous case. We then have for the arm-to-arm separations and phase spreads
\be
\phi_n - \phi_\mathrm{P} = \frac{2\pi n}{m_*} - \int_{r_\mathrm{out}}^r \frac{\Delta k_{m_*}(r^\prime)}{m_*}\mathrm{d}r^\prime,
\ee
and
\be
\delta\phi_n = \left|2\pi n + \int_{r_\mathrm{out}}^r \frac{\Omega^2(r^\prime)\mathrm{d}r^\prime}{c_\mathrm{s}^2(r^\prime)k_{m_*}(r^\prime)}\right|.
\ee

For the case $\omega_\mathrm{p} = 0$, as adopted in \S \ref{sec:passive}, the expressions for $\phi_n - \phi_\mathrm{P}$ and $\delta\phi_n$ take on a simple forms, 
\be
\phi_n - \phi_\mathrm{P} = \frac{2\pi n}{m_*} + \left[1 - \frac{(m_*^2-1)^{1/2}}{m_*}\right] \frac{g(r/r_\mathrm{out})}{h_\mathrm{out}},
\ee
and
\be
\label{eq:dphin_passive}
\delta\phi_n = \left| 2\pi n + \frac{g(r/r_\mathrm{out})}{h_\mathrm{out}(m_*^2-1)^{1/2}} \right|,
\ee
where
\be
g(x) = 
\begin{cases}
\frac{2}{q-1}\left[x^{(q-1)/2} - 1\right] & (q \neq 1), \\
\ln(x) & (q = 1).
\end{cases}
\label{eq:gdef}
\ee

\subsubsection{Inner/Outer Disk Asymmetry}
\label{subsec:theory_inner_outer}

Our numerical results for planet-driven spirals indicate that multiple spirals robustly form in the inner disk, while only a single spiral forms in the outer disk (except for small $h_\mathrm{p}$, for which a weak secondary spiral is present). Evidently, this asymmetry is related to the behavior of the integral in equation~(\ref{eq:arm_width_theory}) for small versus large $r$.

In the outer disk, the integral remains bounded. For $q = 1$, its value for $r \rightarrow \infty$ is exactly
\be
\int_{r_+}^\infty \frac{\Omega^2(r^\prime)\mathrm{d}r^\prime}{c_\mathrm{s}^2(r^\prime)k_m(r^\prime)} = -\frac{2}{3h_\mathrm{p}} \frac{\ln[m-(m^2-1)^{1/2}]}{(m^2-1)^{1/2}}.
\ee
For $q < 1$, the value of the integral is even less than given above, since the integrand is proportional to $r^{(q-6)/2}$ for large $r$. Therefore, for $q=1$ and setting $m\approx (2h_p)^{-1}$ we have
\be
\label{eq:dphin_infty}
\delta\phi_n(\infty) \approx \left|2\pi n - \frac{\pi}{4} - \frac{4}{3}\ln(h_\mathrm{p})\right|.
\ee
Consider $\delta\phi_n$ with $n = -1$. If this quantity becomes sufficiently small in the outer disk, a secondary spiral may form. For the fiducial parameters ($h_\mathrm{p} = 0.1$), $\delta\phi_{-1}(\infty) \approx 3.95$. Since this is larger than the critical phase spread necessary for secondary spiral arm formation given in Table~\ref{tab:dphi_values}, no such spiral forms in the outer disk. 

However, equation~(\ref{eq:dphin_infty}) also indicates that $\delta\phi_{-1}$ may become small enough to produce a secondary spiral if $h_\mathrm{p}$ is small enough. Indeed, in our numerical calculations, a weak secondary spiral is found for $h_\mathrm{p} = 0.05$. But the weak (logarithmic) dependence on $h_\mathrm{p}$ in equation~(\ref{eq:dphin_infty}) indicates that $\delta\phi_{-1}$ can become small enough to produce a {\it strong} secondary spiral only if $h_\mathrm{p}$ is very small.

On the other hand, in the inner disk, for $r \ll r_\mathrm{p}$, one can show that
\be
\int_{r_-}^r \frac{\Omega^2(r^\prime)\mathrm{d}r^\prime}{c_\mathrm{s}^2(r^\prime)k_m(r^\prime)} \approx \frac{g(r/r_\mathrm{p})+\eta}{(m^2-1)^{1/2}h_\mathrm{p}},
\ee
where $\eta$ is an order unity constant. For $q \le 1$, $g(r/r_\mathrm{p}) \gg 1$ for small $r$, and so $\delta\phi_n$ (equation~\ref{eq:arm_width_theory}) must cross zero (indicating constructive interference) before diverging as $r \rightarrow 0$. In other words, it is always possible to find a sufficiently small $r$ such that $\delta\phi_n < \Delta\phi_0$ ($n \ge 1$) for any $\Delta\phi_0$, and so secondary arm formation is unavoidable in the inner disk.


\subsection{Application of Theory to Our Numerical Results}


\begin{figure*}
\begin{center}
\includegraphics[width=0.99\textwidth,clip]{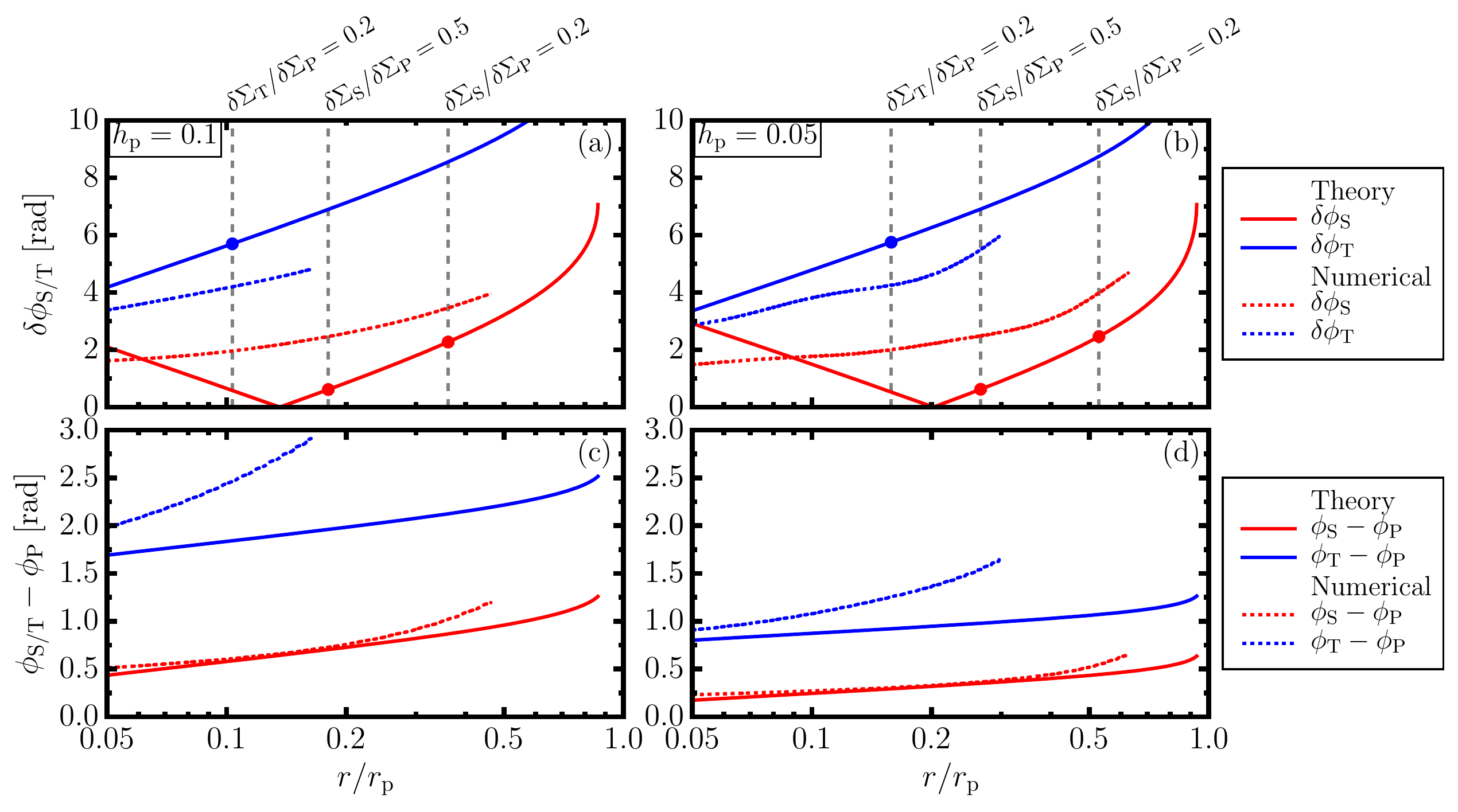}
\caption{The secondary/tertiary phase spreads (equation~\ref{eq:arm_width_theory}; top panels) and positions (equation~\ref{eq:arm_sep_theory}; bottom panels) predicted by our theoretical phase argument (solid curves) and estimated using numerical calculations (dotted curves). Two different cases are shown: $h_\mathrm{p} = 0.1$ (left) and for $h_\mathrm{p} = 0.05$ (right), both with $q = p = 1$. In the top panels, the dashed vertical lines indicate the radii corresponding to several critical relative amplitudes of the secondary/tertiary spirals. The values of the theoretical phase spreads at these critical radii, highlighted by the filled points, are insensitive to the disk parameters. Note that the ``reflection'' feature of the theoretical $\delta\phi_\mathrm{S}$ in panels a--b corresponds to a sign change of the quantity inside the absolute value symbol in equation (\ref{eq:arm_width_theory}). This feature is not seen in the numerical $\delta\phi_\mathrm{S}$, possibly due to the breakdown of the expansion (\ref{eq:dphimn}) when the spread in the mode phases is small, as a result of the discreteness of $m$.}
\label{fig:dphi_phi_theory}
\end{center}
\end{figure*}

We now apply analytical results derived above to understanding certain features of the linear calculations presented in \S \ref{sec:results} and \S \ref{sec:passive}. In Figure~\ref{fig:dphi_phi_theory}a--b we display the theoretical spiral arm phase spreads given by equation~(\ref{eq:arm_width_theory}) for two different values of $h_\mathrm{p}$. The critical radii $r_{\mathrm{S}20}$, $r_{\mathrm{S}50}$, and $r_{\mathrm{T}20}$ found in our linear calculations are also indicated. We see that the values of the relevant $\delta\phi_n$'s at these critical locations are relatively insensitive to $h_\mathrm{p}$ (as well as to the other disk parameters; see Table~\ref{tab:dphi_values}). Therefore, equation~(\ref{eq:dphi_criterion}) with properly calibrated $\Delta\tilde\phi_0$ can be used to reliably predict the location at which the secondary spiral forms. The phase spreads estimated from the numerical spiral arm widths (full width at half maximum for $n$-th arm, FWHM$_n$), $\delta\phi_n = m_*\times \mathrm{FWHM}_n$, which can be compared to the theoretical $\delta\phi_n$'s, are also shown. The theoretical and numerical spiral arm phase spreads show qualitative agreement. They both vary similarly with $r$, although the exact values differ. This may be in part due to the the ambiguity in quantifying the width of a spiral arm found in the numerical calculation (e.g., using the FWHM instead of some other metric). Figure~\ref{fig:dphi_phi_theory}c--d shows the theoretical arm separations (equation~\ref{eq:arm_sep_theory}) as well as the arm separations from our numerical results. The theoretical calculation accurately predicts the azimuthal separation between the primary and secondary spirals, but only roughly predicts the separation of the tertiary and primary spirals. This discrepancy is possibly due to the fact that the tertiary spiral is not dominated by modes with $m \approx m_*$, as assumed in equation~(\ref{eq:arm_sep_theory}), but rather by modes with $m \lesssim m_*$.

The theoretical predictions of the critical radii for secondary spiral formation (for the case of planet-driven spirals) obtained using equations~(\ref{eq:dphi_criterion})--(\ref{eq:arm_width_theory}) are shown in Figure~\ref{fig:secondary_amplitude}. These predictions are calibrated using the values of $\Delta\tilde\phi_0$ found for the fiducial parameters (i.e. for $h_\mathrm{p} = 0.1$). That is, the critical values of $\delta\phi_1$ corresponding to different secondary/primary amplitude ratios given in Table~\ref{tab:dphi_values} for the fiducial parameters are taken to be universal values applicable for all parameters. These predictions based on simple phase arguments give excellent agreement with the numerical results over a range of disk parameters. 

The equivalent predictions for the case of passive spirals are shown in Figure~\ref{fig:secondary_amplitude_homogeneous}. These predictions also agree well with the numerical results. There are, however, some discrepancies for wide spirals (Fig.~\ref{fig:secondary_amplitude_homogeneous}c). This may be due to the fact that wide spirals are dominated by a small number of azimuthal modes, diminishing the accuracy of the approximation (\ref{eq:dphimn}) for the spread in the $\phi_{m,n}$'s.

Given the success of our analytic arguments in explaining some key features of our numerical calculations, we can use them to interpret some features of secondary and higher order spiral arms found in this work. The weak dependence of the characteristics of multiple arm on surface density slope $p$ found in both inhomogeneous (\S \ref{sec:results}) and homogeneous (\S \ref{sec:passive}) cases is easily understood in the context of mode interference. We showed that this process is well described by the WKB approximation, in which the mode phases are independent of the $\Sigma$ profile. Rather, they depend only on $h(r)$, which in our parameterization is fully described by the parameters $h_\mathrm{p}$ and $q$. 

The weak remaining dependence on $p$ in the planet-driven case (see Figure~\ref{fig:secondary_amplitude}c) is likely due to the shift of the Lindblad resonances at which the modes are launched. In the homogeneous case, our choice of $\omega_\mathrm{p} = 0$ effectively places Lindblad resonances far outside the computation domain suppressing any sensitivity to $p$, as stated in \S \ref{sec:passive}.

We can also explain why in the homogeneous case, (azimuthally) narrower patterns of $\delta\Sigma$ produce secondary arm at smaller radii, see Figure~\ref{fig:secondary_amplitude_homogeneous}c. Narrower perturbations have their power concentrated in modes of higher $m$ (i.e, they have a larger values of $m_*\approx (2\sigma)^{-1}$) compared to azimuthally wider perturbations. Equation (\ref{eq:dphin_passive}) then predicts that $\delta \phi_n$ becomes small (resulting in phase coherence for $n=1,2$ and so on) at lower $r/r_{\rm out}$ as $\sigma$ is decreased (and $m_*$ is increased correspondingly); note that $g(x)<0$ in the inner disk, see equation (\ref{eq:gdef}). Therefore, narrower perturbations produce high order spiral arms at smaller radii than the wider ones do.

\begin{table}
\caption{Values of $\delta\phi_n$ at locations corresponding to several relative amplitudes of the secondary/tertiary spiral arm (where for example, $r_{\mathrm{S}50}$ indicates a $50\%$ relative amplitude of the secondary to the primary).}

\begin{tabular}{ccc|ccc}

    \hline

    $h_\mathrm{p}$ & $q$ & $p$ & $\delta\phi_1(r_{\mathrm{S}20})$ & $\delta\phi_1(r_{\mathrm{S}50})$ & $\delta\phi_2(r_{\mathrm{T}20})$ \\ \hline
    
        $0.05$ & $1$     & $1$     & $2.46$ & $0.62$ & $5.75$ \\
        $0.07$ & $1$     & $1$     & $2.39$ & $0.53$ & $5.92$ \\
        $0.10$ & $1$     & $1$     & $2.27$ & $0.62$ & $5.70$ \\
        $0.15$ & $1$     & $1$     & $2.05$ & $0.52$ & $5.08$ \\
        $0.10$ & $1/2$ & $1$     & $2.19$ & $0.44$ & $5.59$ \\
        $0.10$ & $0$     & $1$     & $2.12$ & $0.28$ & $5.50$ \\
        $0.10$ & $1$     & $0$     & $2.12$ & $0.33$ & $5.43$ \\
        $0.10$ & $1$     & $3/2$ & $2.36$ & $0.78$ & $5.85$ \\
      
\end{tabular}
\label{tab:dphi_values}
\end{table}


\section{Discussion}
\label{sec:discussion}


The calculations presented in this work (except for \S \ref{sect:num_val}) are explicitly linear. At the same time, it is well known that nonlinear effects play an important role in the propagation and damping of spiral waves, as well as the evolution of the disk (e.g., \citealt{GR01,R02,R16,AR18}). A planet-driven spiral wake begins to shock at a distance of $L_\mathrm{sh}\sim (M_\mathrm{p}/M_\mathrm{th})^{-2/5} H_p$ from the planet (in the local approximation), evolving into a wide ``N''-shaped wave \citep{GR01}. In the process it deposits its angular momentum into the disk material, so that the angular momentum flux of the wave is no longer conserved (as it is in linear theory). This process is modified by the presence of a secondary spiral \citep{AR18}. Injection of angular momentum originally carried by the spiral wave into the disk material drives the evolution of the disk \citep{AR18} and causes gap opening \citep{R02b}. 

Our linear calculations are strictly valid only as long as the appearance of the secondary spiral is not preceded by the shocking of the primary arm. This condition sets an upper limit on the allowed planet mass. Indeed, if the secondary spiral emerges after the wake travels a distance $\zeta H_p$ in the inner disk, then the condition $L_\mathrm{sh}\gtrsim \zeta H_p$ implies that $M_p\lesssim M_\mathrm{th}\zeta^{-5/2}$. In our $h_\mathrm{p} = 0.1$ calculation the secondary arm forms at $\approx 5H_\mathrm{p}$ interior to the planet, meaning that $\zeta\approx 5$ and $M_p\lesssim 0.02 M_\mathrm{th}$ is needed for our linear calculation to capture the formation of multiple spirals in quantitative detail. However, at the qualitative level our calculation should remain valid at substantially higher values of $M_p$ (e.g., because nonlinear evolution has only a marginal effect on the analytical phase coherence calculation presented in \S \ref{sec:analytic}). 

Nonlinear evolution also affects the morphology of the spirals in the high-$M_p$ regime. Numerical simulations have shown that the azimuthal separation of the secondary and primary arms is $\approx 60^\circ$ for low-mass ($M_\mathrm{p} \ll M_\mathrm{th}$) planets, in agreement with our prediction from linear theory, but increases up to $\approx 180^\circ$ for massive ($M_\mathrm{p} \gg M_\mathrm{th}$) planets \citep{Dong2015,Fung2015}. This transition is caused by the steady azimuthal broadening of the spiral wake due to its nonlinear evolution in the ``N-wave'' regime \citep{GR01,R02,Zhu2015}. Therefore, the secondary spirals which we find to form by linear processes should be regarded just as \textit{precursors} to the fully-fledged secondary spiral arms/shocks (see Section 6 of \citealt{AR18}). In our linear calculations, the tertiary/quaternary arms are always very weak, and so it is unclear how they are affected by nonlinear effects. However note that \citet{Dong2017} reported the presence of these higher-order arms in nonlinear simulations, although found them to be destroyed by moderate viscosity.

We note that the global treatment, i.e., accounting for the cylindrical geometry (as opposed to the local, shearing sheet approximation), is critical for capturing the formation of multiple spiral arms. We find that the distance from the planet at which the secondary arm forms (which we have defined as point at which its amplitude is $10\%$ of the primary), measured in terms of $H_\mathrm{p}$, is a decreasing function of $h_\mathrm{p}$: the secondary arm forms at a distance of $\approx 5H_\mathrm{p}$ from the planet for $h_\mathrm{p} = 0.1$, and at $\approx 7.5H$ from the planet for $h_\mathrm{p} = 0.05$. In the local (shearing sheet/box) approximation, corresponding to the limit $h_\mathrm{p} \rightarrow 0$, in which $H_\mathrm{p}$ is the characteristic length scale, this implies that secondary spirals form at $|r-r_\mathrm{p}|/H_p \rightarrow \infty$. In other words, higher-order spirals would not be captured in the shearing sheet approximation. Furthermore, secondary spirals form almost exclusively in the inner disk, and not in the outer disk, an asymmetry that cannot arise in the shearing sheet framework.


\subsection{Comparison with Other Work}


The numerical calculations presented in this paper largely follow those of \citet{OL02}, although we give a more detailed analysis of the results. However, \citet{OL02} did not report the presence of multiple spirals in their calculations. We find two main reasons for this. First, they only solved for the perturbation structure down to a radius of $0.3 r_\mathrm{p}$ (unlike our calculations, which extend to $0.05 r_\mathrm{p}$, allowing the secondary arm to be fully captured). From our Figure~\ref{fig:arm_profiles_h}, we see that for our fiducial parameters, the secondary arm is still quite weak at that radius, with an amplitude about four times smaller than that of the primary arm. In their Figure 5, the first hint of a secondary arm becomes visible near the inner disk edge, however, it went unnoticed in their discussion. Second, the calculations by \citet{OL02} were restricted to the case of a constant disk aspect ratio, i.e., $q = 1$ in our notation. For flared disks, with $q < 1$, the secondary arm emerges, and also overtakes the primary in amplitude closer to the planet (see Figs.~\ref{fig:arm_profiles_q} and \ref{fig:secondary_amplitude}), making its presence more apparent. Finally, we note that \citet{OL02} did point out that the phases $\psi_m$ of modes with different $m$ eventually diverge (logarithmically for the case $q = 1$) towards the inner disk, so that their constructive interference fails resulting in the partial dissolution of the primary arm. However, they missed the fact that the same process also results in convergence of $\psi_m$ to an integer multiple of $2\pi$ (implying the same value of $\delta\Sigma_m$) at different azimuthal locations in the disk, giving rise to higher order spirals (see \S \ref{sec:analytic}).

\citet{R02} arrived at the one-armed spiral solution using a method different from \citet{OL02}. In his case the inability to capture the formation of higher-order arms is likely caused by a certain assumption used in the derivation of the linear wake shape, namely the conservation of the Riemann invariant along the characteristics that cross (rather than follow) the wake. Small changes of this invariant at the wake crossings, neglected in \citet{GR01} and \citet{R02}, could be responsible for the eventual emergence of the secondary spiral. This conjecture is supported by the fact that secondary spiral emerges closer to the planet in disks with lower $h_\mathrm{p}$: the number of wake crossings by characteristics (per fixed radial interval) grows as $h_\mathrm{p}$ goes down, facilitating breakdown of the one-spiral solution.

Some other ideas for the origin of secondary spirals have been advanced, in particular, nonlinear effects related to ultraharmonic resonances with the planet \citep{Fung2015}. We do not find these explanations persuasive as, first, we reproduce multiple spirals in the framework of a purely linear calculation. Second, our calculations of passive propagation of a wake with $\omega_\mathrm{p} = 0$ in \S \ref{sec:passive} do not feature any resonances, and yet, they do result in secondary spirals.

\subsection{Applications}

A possible connection between the multiple spirals driven by a planet and the multiple gaps and rings seen in some protoplanetary disks was suggested by \citet{DongGaps2017,DongGaps2018} and \citet{Bae17}. In the picture put forth by these authors, a low-mass planet in a low-viscosity disk produces multiple spiral arms, each of which shock, dissipate, and open a gap at some distance from the planet. The location of the secondary gap, attributed to the dissipation of the secondary spiral, was given by \citet{DongGaps2018} as a function of planet mass and disk thickness. Our linear prediction for the location at which the secondary spiral forms is exterior to their predicted gap location for small planet masses ($M_\mathrm{p} \lesssim 0.2 M_\mathrm{th}$). This is consistent with the scenario in which the secondary spiral first forms in a linear fashion at some distance from the planet, then propagates inwards, evolving nonlinearly, before shocking and opening a gap. For larger planet masses, \citet{DongGaps2018} predict a secondary gap at a location too close to the planet for a secondary spiral to have formed in linear theory. In this case, nonlinear effects clearly play a role not just in the dissipation of the spiral, but also in its formation.

The multiple spiral features observed in some protoplanetary disks may be produced by planets \citep{Dong2015}. The $180^\circ$ separation of these spirals requires massive planets to produce, and so nonlinear effects cannot be neglected in these cases. We nonetheless expect the linear mechanism described in this work to play an important role in providing the conditions necessary for the formation of these structures \citep{AR18}.

However, planets are not the only possibility. As we showed in \S \ref{sec:passive}, any spiral arm, regardless of its origin, inevitably evolves into multiple spirals as it propagates through the differentially rotating disk (note that the assumption of a Keplerian profile for $\Omega(r)$ is not essential for the arguments advanced in \S \ref{sec:analytic}). Therefore, any mechanism capable of producing at least one spiral arm necessarily produces multiple spiral arms. Possible mechanisms include gravitational instability (e.g., \citealt{DongGI2015}), accretion from an infalling envelope \citep{Lesur2015,Hennebelle2017}, shadows/non-axisymmetric illumination \citep{Montesinos2016}, and vortices \citep{Paardekooper2010}. We only require that density waves with a range of azimuthal mode numbers are excited and that they are at least somewhat in phase with one another, so that one or a few well-defined spiral arms (rather than many flocculent spirals) are produced.


\section{Summary}
\label{sec:sum}


We explored the origin of multiple spiral arms, which are often observed in protoplanetary disks and also found in numerical simulations of disks with massive perturbers. The two-dimensional structure of surface density perturbations induced by a planet (as well as that of a passive spiral) was computed using linear theory of density wave excitation and propagation \citep{GT79} in the low planet mass (low amplitude) regime. 

We find that, in addition to the strong single spiral arm excited by the planet in agreement with past studies \citep{OL02,R02}, a secondary spiral arm (and often a tertiary arm, quaternary arm, and so on) robustly forms in inner disk in the linear regime. The secondary arm first appears at about $r = (0.4 - 0.6) r_\mathrm{p}$, and, though initially weak, becomes stronger and narrower towards the center of the disk, eventually exceeding strength of primary arm at $\approx 0.1 r_\mathrm{p}$. As the primary arm propagates into the inner disk, the angular momentum flux it carries gets steadily transferred to these higher-order spiral arms. In the outer disk, we find that a secondary spiral arm typically does not form, except for the coldest disk we considered, with $h_\mathrm{p} = 0.05$. 

We provide analytical arguments extending the reasoning of \citet{OL02}, which show that secondary spiral arms form as a result of the constructive interference among different azimuthal modes in the inner disk. Our treatment, which implicitly takes into account the global variation of both the phases and amplitudes of different linear modes in a self-consistent manner, thus corroborates the semi-quantitative, phase coherence picture previously put forth by \citet{BZ18a} in the WKB limit. The gravitational potential of the planet does not play a role in this process. Rather, the planet only seeds the initial perturbation, which then propagates passively and spawns higher-order spiral arms. This is confirmed by the persistent emergence of multiple spirals also in our linear calculations of the passive inward propagation of an imposed spiral wake, free from the influence of an external potential (following the setup of \citealt{AR18}). 

Our results clearly demonstrate that the formation of secondary spirals is an intrinsically linear process, which serves as a precursor for subsequent nonlinear evolution resulting in a formation of multiple shocks in the disk. These calculations should help us better understand planet-driven evolution of protoplanetary disk \citep{GR01,R02}, including the formation of multiple gaps in such disks \citep{Bae17,DongGaps2017,DongGaps2018}. We use them to understand the details of the global distribution of the torque exerted by an embedded planet on a disk in Miranda \& Rafikov (in prep.).

\acknowledgements 

Financial support for this study has been provided by NSF via grant AST-1409524 and NASA via grant 15-XRP15-2-0139. We thank Stephen Lubow and Gordon Ogilvie for valuable comments. We are grateful to Wing-Kit Lee for a careful reading of this paper and for a number of useful suggestions.


\appendix


\section{Numerical Procedure}
\label{app:solution_method}


\begin{figure}
\begin{center}
\includegraphics[width=0.49\textwidth,clip]{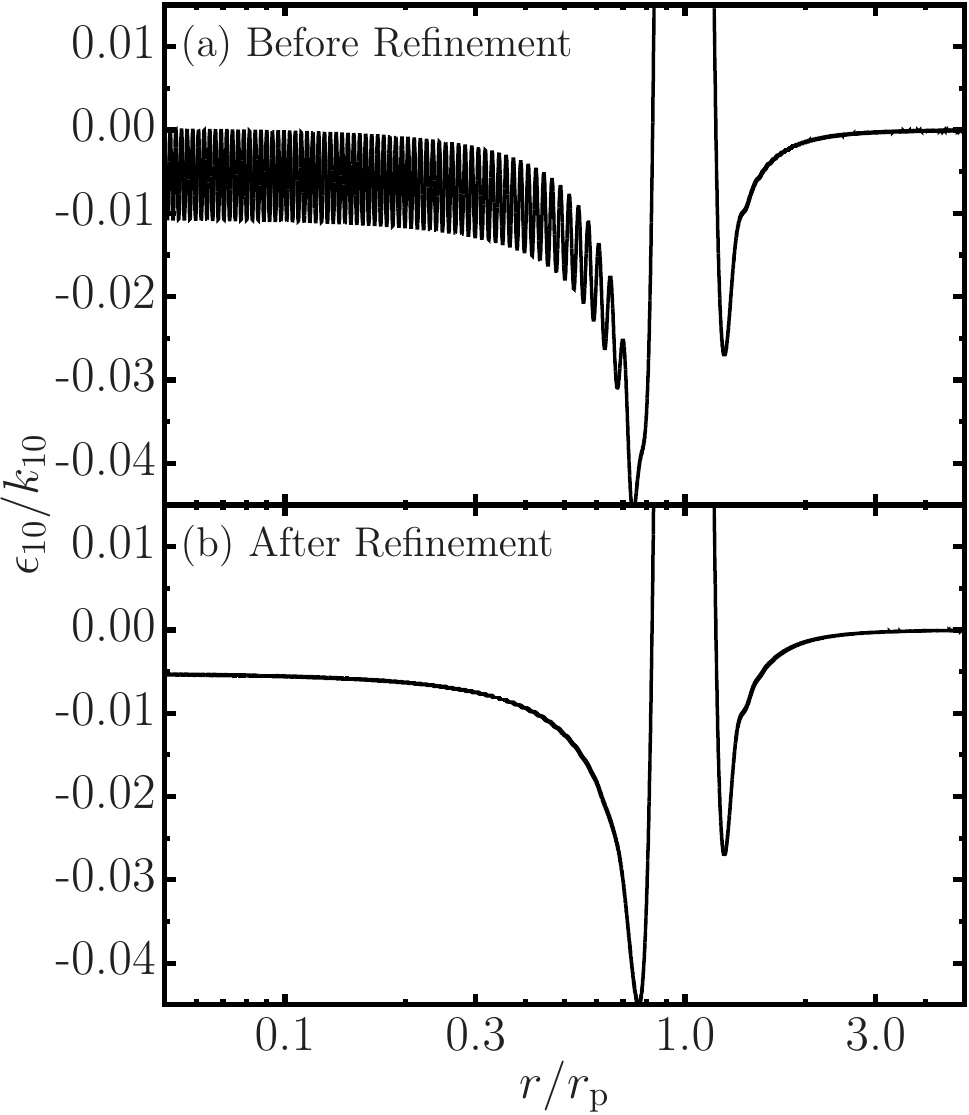}
\caption{Fractional error in the phase gradient before and after refinement for the $m = 10$ mode, with the fiducial parameters $h_\mathrm{p} = 0.1$, $q = 1$ and $p = 1$.}
\label{fig:phase_refine}
\end{center}
\end{figure}

\begin{figure*}
\begin{center}
\includegraphics[width=0.99\textwidth,clip]{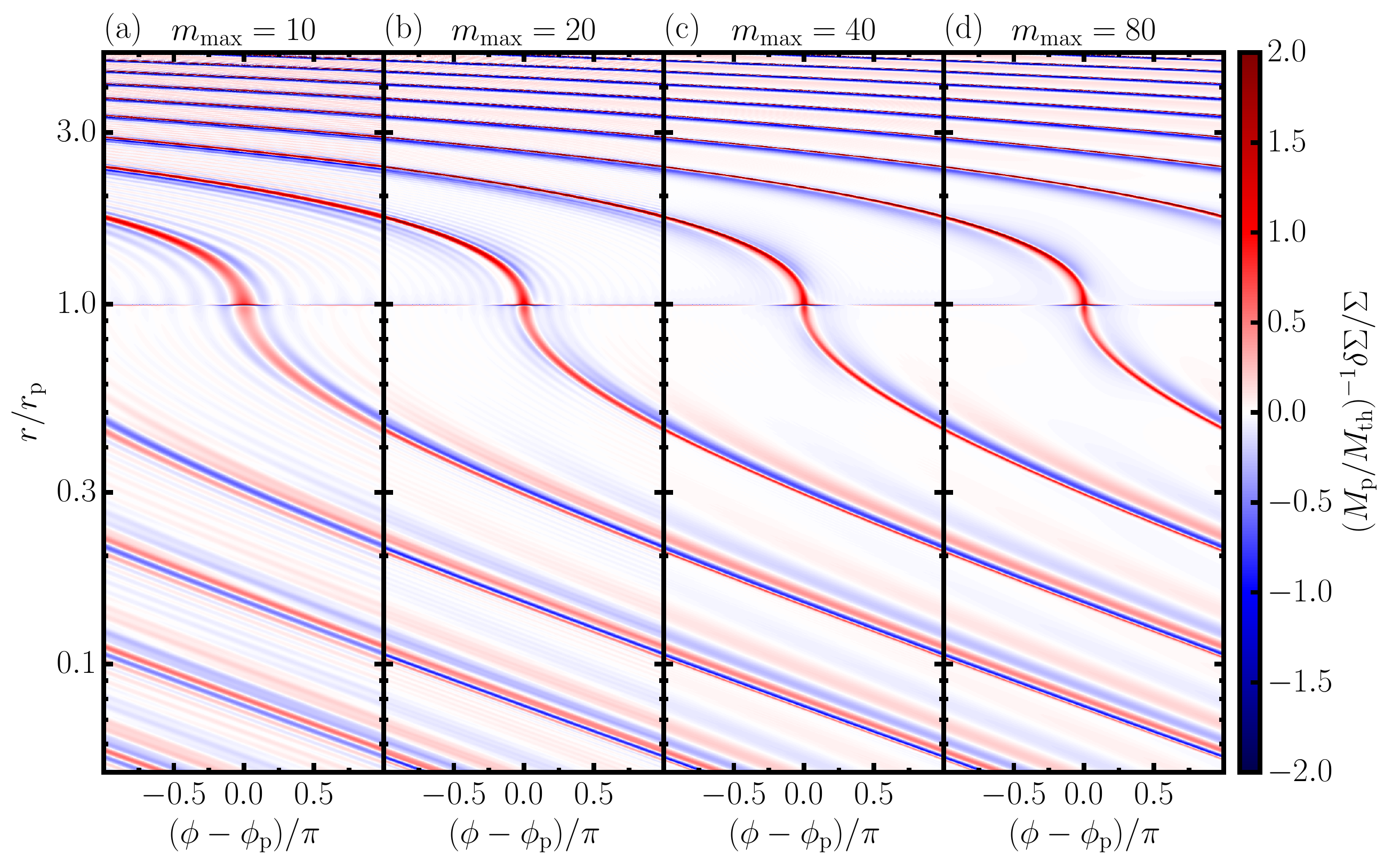}
\caption{Two-dimensional map in polar coordinates $(r,\phi)$ of the fractional surface density perturbation $\delta\Sigma/\Sigma$ (scaled by the ratio of the planet mass, $M_\mathrm{p}$, to the thermal mass, $M_\mathrm{th} = h_\mathrm{p}^3 M_*$), produced using different values of $m_\mathrm{max}$, the number of azimuthal modes synthesized. The case shown here uses the fiducial parameters.}
\label{fig:m_max}
\end{center}
\end{figure*}

\subsection{Mode Solutions}

Here we give a detailed description of the numerical method used for producing solutions of equation~(\ref{eq:master_eqn}). We compute the values of Laplace coefficients, as well as their derivatives required for equation~(\ref{eq:master_eqn}) (by first differentiating the integrand of equation~(\ref{eq:laplace_coef}) with respect to $\alpha$) using numerical quadrature. We remove the corotation pole ($\tilde{\omega} = 0$) in equation~(\ref{eq:master_eqn}) by replacing $\tilde{\omega}$ with $\tilde{\omega} + \mathrm{i}\delta$, where $\delta$ is a small positive real constant (KP93). We choose $\delta = 10^{-6}$.

The solution method closely follows the technique described in detail by KP93, with several small differences. We first produce two linearly independent solutions, $\delta h_\mathrm{H}^{(1)}$ and $\delta h_\mathrm{H}^{(2)}$ of the homogeneous version of equation (\ref{eq:master_eqn}), by integrating outwards from corotation starting from arbitrary initial values, as well as an inhomogeneous solution, $\delta h_\mathrm{IH}$. The desired solution, which satisfies any pair of specified boundary conditions at the boundaries $r_\mathrm{in}$ and $r_\mathrm{out}$, can be expressed as
\be
\label{eq:soln_lincomb}
\delta h_m = a_1 \delta h_\mathrm{H}^{(1)} + a_2 \delta h_\mathrm{H}^{(2)} + \delta h_\mathrm{IH},
\ee
where $a_1$ and $a_2$ are constants determined by the specific form of the boundary conditions. We choose outgoing wave boundary conditions, with
\be
\delta h_m^\prime(r_\mathrm{in}) = C_\mathrm{in} \delta h_m(r_\mathrm{in}), \ \delta h_m^\prime(r_\mathrm{out}) = C_\mathrm{out} \delta h_m(r_\mathrm{out}),
\ee
where
\be
\label{eq:outgoing}
C = \mathrm{i}k + \frac{1}{2}\frac{\mathrm{d}}{\mathrm{d}r}\ln\left(\frac{D_S}{r\Sigma k}\right),
\ee
representing an outgoing wave in the WKB limit (e.g., \citealt{Tsang2008}). Here $k = (-D_S)^{1/2}/c_\mathrm{s}$ is the radial wavenumber. Note that second term in equation (\ref{eq:outgoing}), describing the slow change in amplitude of the wave, was not included by KP93 at this step (although they took it into account in an approximate fashion at a later step). Now that we have specified the boundary conditions, we have a system of equations which can be solved for $a_1$ and $a_2$:
\be
\label{eq:bc_sys_1}
\begin{aligned}
& \left[\frac{\mathrm{d}}{\mathrm{d}r}\delta h_\mathrm{H}^{(1)}(r_\mathrm{in}) - C_\mathrm{in} \delta h_\mathrm{H}^{(1)}(r_\mathrm{in})\right] a_1 \\
+ & \left[\frac{\mathrm{d}}{\mathrm{d}r}\delta h_\mathrm{H}^{(2)}(r_\mathrm{in}) - C_\mathrm{in} \delta h_\mathrm{H}^{(2)}(r_\mathrm{in})\right] a_2 \\
= & C_\mathrm{in} \delta h_\mathrm{IH}(r_\mathrm{in}) - \frac{\mathrm{d}}{\mathrm{d}r} \delta h_\mathrm{IH}(r_\mathrm{in}),
\end{aligned}
\ee
\be
\label{eq:bc_sys_2}
\begin{aligned}
& \left[\frac{\mathrm{d}}{\mathrm{d}r}\delta h_\mathrm{H}^{(1)}(r_\mathrm{out}) - C_\mathrm{out} \delta h_\mathrm{H}^{(1)}(r_\mathrm{out})\right] a_1 \\
+ & \left[\frac{\mathrm{d}}{\mathrm{d}r}\delta h_\mathrm{H}^{(2)}(r_\mathrm{out}) - C_\mathrm{out} \delta h_\mathrm{H}^{(2)}(r_\mathrm{out})\right] a_2 \\
= & C_\mathrm{out} \delta h_\mathrm{IH}(r_\mathrm{out}) - \frac{\mathrm{d}}{\mathrm{d}r} \delta h_\mathrm{IH}(r_\mathrm{out}).
\end{aligned}
\ee

As in KP93, once we have found the values of $a_1$ and $a_2$ and constructed a solution satisfying the boundary conditions, we generate a new inhomogeneous solution $\delta h_\mathrm{IH}$ using the ``correct'' values of $\delta h$ and $\delta h^\prime$ at corotation (i.e., the ones found from the previously constructed solution). The system of equations (\ref{eq:bc_sys_1}--\ref{eq:bc_sys_2}) is then re-solved for new values of $a_1$ and $a_2$. This process ensures that the final solution is robust with respect to the specific choice of (arbitrary) homogeneous solutions used in its construction (equation~\ref{eq:soln_lincomb}), because it tends to reduce the absolute values of $a_1$ and $a_2$, so that the contributions to the correct solution from the homogeneous solutions are small.

As a final refinement step, also described in KP93, we minimize the amplitude of the oscillations of the ``phase gradient error'' near the boundaries. The phase gradient error, defined as
\be
\epsilon = \mathrm{Arg}(\delta h)^\prime - k,
\ee
serves as a diagnostic of how close the solution is to an outgoing WKB wave. The phase gradient error of the initial solution exhibits oscillations, representing contamination by an in incoming wave, near the boundaries. We seek to minimize the amplitude of these oscillations by slightly adjusting the constants $C_\mathrm{in}$ and $C_\mathrm {out}$ (so that they are no longer given exactly by equation \ref{eq:outgoing}) characterizing the boundary conditions. In principle there are a variety of ways to achieve this (note that the exact procedure used by KP93 is not specified). We choose to minimize the peak-to-peak amplitudes of the phase gradient error $\epsilon$ within one local scale height of each boundary:
\be
X_\mathrm{in} = [\max(\epsilon) - \min(\epsilon)]_{r_\mathrm{in}<r<r_\mathrm{in}+H(r_\mathrm{in})},
\ee
\be
X_\mathrm{out} = [\max(\epsilon) - \min(\epsilon)]_{r_\mathrm{out}-H(r_\mathrm{out})<r<r_\mathrm{out}}.
\ee
Note that we also tried minimizing the derivative of the phase gradient error $\epsilon^\prime$ at $r_\mathrm{in}$ and $r_\mathrm{out}$, but found that this was not as effective at getting rid of the oscillations. We numerically compute the Jacobian describing the derivatives of $\mathbf{X} = (X_\mathrm{in}, X_\mathrm{out})$ with respect to $\mathbf{C} = (C_\mathrm{in}, C_\mathrm{out})$, and use its inverse to perform one step of the secant method to find the root of $\mathbf{X}(\mathbf{C})$. This greatly reduces the amplitude of the oscillations of the phase gradient error, see Figure~\ref{fig:phase_refine} for an example of this process. In practice this refinement only requires changing the values of $C_\mathrm{in}$ and $C_\mathrm{out}$ by a very small amount ($\lesssim 1\%$). Also note that this refinement produces only very imperceptible changes in the form of $\delta h(r)$ and $\delta h^\prime(r)$, but is potentially important for ensuring that different modes have the correct phase when the interference of many modes is considered.

As noted in \S \ref{sec:setup}, we do not include the indirect potential term $\delta_{m,1}GM_\mathrm{p}r/r_\mathrm{p}^2$ in our calculations. This term is proportional to $r$, and so becomes large for $r \gg r_\mathrm{p}$, in contrast to the direct terms (equation~\ref{eq:potential_components}), which become small for $r \gg r_\mathrm{p}$ (note that they both become small for $r \ll r_\mathrm{p}$). Therefore, for $m \neq 1$, equation~(\ref{eq:master_eqn}) becomes effectively homogeneous for $r \gg r_\mathrm{p}$. Our solution method exploits this fact by using knowledge of the asymptotic behavior of the homogeneous equation to set the outer disk boundary condition (and similarly for the inner boundary condition). However, for $m = 1$, when the indirect term is included, its anomalous behavior at large $r$ calls into question the validity of setting the outer boundary condition in this way. We nonetheless carried out several tests in which the indirect potential was included and the outer boundary condition was set under the homogeneous assumption. We verified that including the the indirect potential in this way only slightly modifies the profile of the density wake in the outer disk, but does not otherwise affect our main results.

\subsection{Mode Synthesis}

The two-dimensional surface density perturbation $\delta\Sigma(r,\phi)$ is synthesized from the mode solutions $\delta h_m$ using equations (\ref{eq:dsigma_dh}) and (\ref{eq:real}). In order to produce an accurate solution, a sufficient number of modes (up to some  $m_\mathrm{max}$) must be used. The solution must be converged with respect to $m_\mathrm{max}$, i.e., the perturbation structure should not change as more modes are added. The value $m_\mathrm{max}$ required for convergence of the two-dimensional surface density is several times larger than the cutoff parameter $m_\mathrm{cut} \approx  h_\mathrm{p}^{-1}$. Figure~\ref{fig:m_max} illustrates this point by revealing spurious features in the distribution of $\delta\Sigma(r,\phi)$ for low $m_{\rm max}=(1-2)h_\mathrm{p}^{-1}$ (panels a--b). Therefore, in all of our calculations we choose $m_\mathrm{max} \approx 8 m_\mathrm{cut}$, for which Figure \ref{fig:m_max}d demonstrates convergence.

\bibliographystyle{apj}
\bibliography{references}

\end{document}